\documentclass[12pt]{article}

\usepackage{epsfig}
\usepackage{comment}
\usepackage{latexsym}
\usepackage{color}
\newcommand{\mysquare}[0]{\raise-.2ex\hbox{{\Large$\Box$}}}
\def\lsim{\mathrel{\rlap {\raise.5ex\hbox{$ < $}}
{\lower.5ex\hbox{$\sim$}}}}
\def\gsim{\mathrel{\rlap {\raise.5ex\hbox{$ > $}}
{\lower.5ex\hbox{$\sim$}}}} \topmargin -1.5cm \textheight=22.5cm
\catcode`\@=11
\newcount\hour
\newcount\minute
\newtoks\amorpm
\hour=\time\divide\hour by60 \minute=\time{\multiply\hour by60
\global\advance\minute by-\hour}
\edef\standardtime{{\ifnum\hour<12 \global\amorpm={am}%
        \else\global\amorpm={pm}\advance\hour by-12 \fi
        \ifnum\hour=0 \hour=12 \fi
        \number\hour:\ifnum\minute<10 0\fi\number\minute\the\amorpm}}
\edef\militarytime{\number\hour:\ifnum\minute<10 0\fi\number\minute}
\def\draftlabel#1{{\@bsphack\if@filesw {\let\thepage\relax
   \xdef\@gtempa{\write\@auxout{\string
      \newlabel{#1}{{\@currentlabel}{\thepage}}}}}\@gtempa
   \if@nobreak \ifvmode\nobreak\fi\fi\fi\@esphack}
        \gdef\@eqnlabel{#1}}
\def\@eqnlabel{}
\def\@vacuum{}
\def\draftmarginnote#1{\marginpar{\raggedright\scriptsize\tt#1}}
\def\draft{\oddsidemargin -.2truein
        \def\@oddfoot{\sl preliminary draft \hfil
        \rm\thepage\hfil\sl\today\quad\militarytime}
        \let\@evenfoot\@oddfoot \overfullrule 3pt
        \let\label=\draftlabel
        \let\marginnote=\draftmarginnote
   \def\@eqnnum{(\theequation)\rlap{\kern\marginparsep\tt\@eqnlabel}%
\global\let\@eqnlabel\@vacuum}  }

\newcommand{\be}[0]{\begin{equation}}
\newcommand{\ee}[0]{\end{equation}}
\newcommand{\ba}[0]{\begin{eqnarray}}
\newcommand{\ea}[0]{\end{eqnarray}}
\def\bs{\begin{subequations}}
\def\es{\end{subequations}}
\def\d{\partial}

\def\thebibliography#1{%
\vskip 0.5cm \centerline{\bf \Large References}
\list{%
[\arabic{enumi}]}{\settowidth\labelwidth{[#1]}
\leftmargin\labelwidth \advance\leftmargin\labelsep
\usecounter{enumi}}
\def\newblock{\hskip .11em plus .33em minus .07em}
\sloppy\clubpenalty4000\widowpenalty4000 \sfcode`\.=1000\relax}

\renewcommand{\theequation}{\arabic{section}.\arabic{equation}}

\renewcommand{\section}{\setcounter{equation}{0}\@startsection
{section}{1}{0mm}{-\baselineskip}{0.5\baselineskip}
{\normalfont\Large\bfseries}}
\renewcommand{\subsection}{\@startsection
{subsection}{2}{0mm}{-\baselineskip}{0.5\baselineskip}
{\normalfont\large\bfseries}}
\renewcommand{\subsubsection}{\@startsection
{subsubsection}{3}{0mm}{-\baselineskip}{0.5\baselineskip}
{\normalfont\normalsize\slshape}}

\usepackage{amssymb,amsfonts}
\usepackage{graphicx}
\usepackage{cite}

\def\bc{\begin{center}}
\def\ec{\end{center}}
\def\bea{\begin{eqnarray}}
\def\eea{\end{eqnarray}}

\def\s{\sigma}
\def\c{\chi}

\def\d{\delta}
\def\t{\tau}
\def\l{\lambda}
\def\L{\Lambda}

\def\k{\kappa}

\def\ve{\varepsilon}
\def\O{\Omega}
\def\D{\Delta}
\def\taut{\tilde \tau}
\def\P{{\cal P}}
\def\A{{\cal A}}
\def\ad{\dot a}

\def\tt{\tilde t}
\def\dc{\d_\chi^2}
\def\dt{\d'^2_T}

\newcommand{\ie}{{\em i.e. }}

\newcommand{\intN}{\mathbb{N}}
\newcommand{\R}{\mathbb{R}}

\newcommand{\N}{{\cal N}}

\begin{document}

\begin{titlepage}
\begin{flushright}
LPTENS-07/21\\
CPHT-RR024.0407\\
May 2007
\end{flushright}

\vspace{2mm}

\begin{centering}
{\bf \huge Instanton Transition in Thermal and}\\
\vspace{2mm}
{\bf \huge Moduli deformed de Sitter}\\
\vspace{2mm}
{\bf \huge Cosmology$^\ast$}\\

\vspace{6mm}
 {\Large Costas~Kounnas$^{1}$
and Herv\'e~Partouche$^2$}

\vspace{4mm}

$^1$ Laboratoire de Physique Th\'eorique,
Ecole Normale Sup\'erieure,$^\dagger$ \\
24 rue Lhomond, F--75231 Paris cedex 05, France\\
{\em Costas.Kounnas@lpt.ens.fr}

\vskip .1cm

$^2$ {Centre de Physique Th\'eorique, Ecole
Polytechnique,$^\diamond$
\\
F--91128 Palaiseau cedex, France\\
{\em Herve.Partouche@cpht.polytechnique.fr}}

\vspace{6mm}

{\bf\Large Abstract}

\end{centering}

\vspace{0mm}

\begin{quote}
We consider the de Sitter cosmology deformed by the presence of a thermal bath
of radiation and/or time-dependent moduli fields. Depending on the parameters, either a first or second order phase transition can occur. 

In the first case, an instanton  
allows a double analytic continuation. It induces a probability to enter the inflationary evolution by tunnel effect from another cosmological
solution. The latter starts with a big bang and, in the case the transition
does not occur, ends with a big crunch. A temperature duality exchanges the
two cosmological branches. In the limit where the pure de Sitter universe is
recovered,  the tunnel effect reduces to a ``creation from nothing'', due to
the vanishing of the big bang branch. However, the latter may be viable in some range of the deformation parameter. In the second case, there is a smooth evolution from a big bang to the inflationary phase. 
\end{quote}

\vspace{5pt} \vfill
\hrule width 6.7cm \vskip.1mm{\small \small \small \noindent
$^\ast$\ Research partially supported by the EU (under the contracts MRTN-CT-2004-005104, MRTN-CT-2004-512194, MRTN-CT-2004-503369, MEXT-CT-2003-509661), INTAS grant 03-51-6346, CNRS PICS 2530 and 3059,  and ANR (CNRS-USAR) contract  05-BLAN-0079-01.\\
$^\dagger$\ Unit{\'e} mixte  du CNRS et de l'Ecole
Normale Sup{\'e}rieure associ\'ee \`a l'Universit\' e Pierre et Marie Curie (Paris 6), UMR 8549.\\
$^\diamond$\  Unit{\'e} mixte  du CNRS et de  l'Ecole Polytechnique,
UMR 7644.}
\end{titlepage}
\newpage
\setcounter{footnote}{0}
\renewcommand{\thefootnote}{\arabic{footnote}}

\setlength{\baselineskip}{.7cm} \setlength{\parskip}{.2cm}


\section{Introduction}

The recent astrophysical observations indicate that our universe is
in a phase of classical expansion, with a small  positive
cosmological term, $\Lambda$, representing $60-70\%$ of
the total energy density. However, at early time, quantum corrections to this trajectory are
expected to be of first significance. For instance, drastic non perturbative
effects could occur as topology changes, as can be seen in
the context of field theory in semi classical approximation. As an 
example, a topology change is provided by the instability of the five
dimensional Kaluza-Klein (KK) space-time \cite{Witten:1981gj}. In that
case, one 
observes the nucleation of a finite size ``bubble of nothing'' that is
then growing up at the speed of light. The transition is 
described by tunnel effect in terms of an instanton configuration
allowing a ``double analytic continuation''. This means that at two
different 
Euclidean times $\t_{i}$ and $\t_{f}$, analytic continuations to real
times $\t =\t_{i}+it$ and $\t =\t_{f}+it$ are allowed. To be specific,
the configuration admits a time independent  asymptotic behavior
allowing a continuation at $\t_i = + \infty$ to the KK universe, while
another continuation at $\t_f=0$ describes the evolution of a
bubble. 

Actually, the 
KK universe is suffering from a first order phase transition
 where bubbles appear instantaneously, grow and coalesce, so that the space    
``evaporates'' into nothing (after an infinite time, since the volume is
itself infinite). In some sense, some reversed ideas can be
invoked in another example of topology change, namely in Vilenkin's
scenario of the de Sitter space ``creation from  nothing'' \cite{Vilenkin:1982de, Vilenkin:1983xq}. In that case, a finite radius $S^3$ space appears 
instanteously by tunnel effect from a space-time state that amounts
to the empty set. This $S^3$ ``bubble'' is then 
following a de Sitter growing up evolution. The transition involves an
instanton, whose shape is an $S^4$ hemisphere. An analytic continuation
on its boundary amounts to ``gluing'' a de Sitter universe, while the
fact that the instanton is compact with no other boundary to
analytically continue allows an
interpretation in terms of a transition from an empty set.  

Concerning the birth of the de Sitter space, the instanton method
provides an estimate of the transition probability equivalent to the
one derived from the Hartle-Hawking wave function  \cite{Hartle:1983ai} approach. The transition 
amplitude and wave function $\Psi$ being proportional to $e^{-S_E}$,
where $S_E$ is the 
Euclidean action, the probability of the event is  
\begin{equation}
p\propto e^{-2S_E}\, .
\end{equation}
A selection principle  of the cosmological constant $\L$ based on the
extremum of $p$  leads to a favored value $\L\to0_+$, \cite{Hawking:1984hk}. 
However, a present too small cosmological term cannot account for the $60\%$ to $70\%$ of the total energy density, which is necessary to explain the recent dark energy data. 
To remedy to this
problem, it as been stressed \cite{Sarangi:2005cs, Sarangi:2006eb} that since the de Sitter space
has an horizon, it can be associated a Hawking temperature. Thus,
quantum fluctuations of the metric (or any massless mode introduced in
the model) induce a space filling thermal bath. The latter implies an
additional radiation term in the action\footnote{See \cite{Watson:2006px} for a cosmological scenario based on a cascade of transitions between vacuum and radiation energy.} and thus a back reaction on the
metric background that deforms the de Sitter solution (see also \cite{Brout:1989pw}). In that case, the modified probability transition
derived from the Euclidean action is maximal for a non vanishing
finite value of $\L$. This computation has also been addressed by the
authors of \cite{Brustein:2005yn} and refined in \cite{BouhmadiLopez:2006pf}, who considered the Wheeler-de
Witt equation in the WKB approximation. They found that the tunnel
effect  is not connecting the deformed de Sitter expansion to
``nothing'', but to what they called a thermally excited era. However,
the latter is $\L$-dependent. This point makes the difference with the pure de Sitter space created from a $\L$-independent state, namely the empty set, so that extremizing the transition amplitude with respect to $\L$ was making sense. In the case of the deformed solution, applying this selection rule for $\L$ is thus questionable. 

Our first aim is to reconsider the thermally deformed de Sitter space-time. The previous analysis that involves an instanton transition occurs when the radiation energy density is below some critical value. This case corresponds to a first order phase transition. A double analytic continuation on the Euclidean 
solution implies a probability  to connect two cosmological behaviors
in real time. The first one is the de Sitter like evolution, while the
second one is an era that begins  with a big bang and ends with a big
crunch, when the transition does not occur. If instead the radiation energy density is above the critical value, the phase transition becomes second order. It describes a smooth evolution from a big bang to a de Sitter like behavior.  In all these solutions, the temperature is formally infinite when the radius of the universe vanishes. This means that
these evolutions should be trusted as soon as the temperature passes
below some upper bound such as the string or Hagedorn temperature $T_H$ of
an underlying fundamental string theory. 

We then generalize the model by including time dependent moduli-like fields that imply a further deformation of the de Sitter solution. Since, these deformations respect the Robertson-Walker (RW) isometries, they can be expressed in terms of the scale factor $a$ of the universe. The Hubble parameter is taking the form 
\begin{equation}
 \label{H}
3\left( {\dot{a} \over a} \right)^2=3\l - { 3k\over a^2}+{ C_R \over a^4} + { C_M \over  a^6} \, ,
\end{equation}
where the constant and $1/a^2$ terms are associated to the cosmological constant and uniform space curvature. The  $1/a^4$ monomial accounts for the 
effective radiation term discussed before, while the on shell moduli
contribution reaches the $1/a^6$ contribution. What we find is that the moduli deformation can 
be fully absorbed in a redefinition of time. We then describe different
cosmological scenarios that depend on $C_R$ and $C_M$. These parameters can be expressed in terms of the cosmological constant, the number of effective ``massless'' degrees of freedom, the temperature at the transition and the slope of the moduli dependence. If this two parameters model is interesting by itself from a formal point of view, its study is also motivated by the fact that it describes more complicated systems once they are considered on shell. For instance, cases involving a scalar field with a non trivial potential and coupled to moduli with non trivial kinetic terms are treated in \cite{KP}.  


\section{Effective cosmological models in minisuperspace}
\label{tunneling}

We focus on models where space is
3-dimensional\footnote{By this, we mean that if the fundamental theory
  is for instance string (or M-)theory, there are compact dimensions
  whose sizes are dynamically constrained to remain of order $l_s$,
  the string length.} and closed.  
The number of degrees of freedom is important, and actually
infinite in the context of string or M-theory. However, a subset of them to be considered in an effective field theory is determined by the ultraviolet cut off set by the fundamental  theory. Some of the simplest
class of 
models that can be constructed are considering the so called
``minisuperspace" (MSS) approximation \cite{Hartle:1983ai, Hawking:1983hj}. They
involve isotropic and homogeneous spaces only, so that the only
gravitationnal degree of freedom is the scale
factor of the universe, to which some matter fields can be coupled
to. The space-time metric is thus  
\begin{equation}
\label{metric}
ds^2=\sigma^2 \left(-N^2(t)dt^2+a^2(t)d\Omega_3^2\right)\; , \quad \mbox{where}\quad \sigma^2={2\over 3\pi M_p^2}\, ,
\end{equation}
$M_p$ is the Planck mass and $\sigma a(t)$ is the radius of a 
3-sphere. $N$ is a gauge dependent function that can be arbitrarily chosen by a redefinition of time\footnote{\label{fo}We shall use in the following the time variable
  $t$ either when the function $N(t)$ remains unspecified, or when it
  is fixed to 
  $N(t)\equiv 1$. We shall instead use another variable, $\tt$, when
  we choose to fix  $N(\tt)=1/\left(1+\l^{-1}\k/a^2(\tt)\right)$, where $\k$ is a constant.}. (In our conventions, $N$ and $t$ are dimensionless.)
In order to include in the effective action the quantum fluctuations of the full
metric  and matter  degrees of freedom of the theory,  we switch  on  a thermal bath which is created simultaneously with the space-time.  The thermal bath involves the degrees of freedom below the temperature  scale $T$ and is consisting in an extra radiation term in the effective MSS action. This amounts to take into account the back reaction of the thermal bath on the space-metric \ie deform the de Sitter solution \cite{Sarangi:2005cs,  Sarangi:2006eb}. 
 
The starting point is a gravity action coupled to some matter fields, whose Lagrangian density is $L_m$ : 
\begin{equation}
\label{action}
S\equiv S_{g}+S_{m}= {M^2_p\over 16 \pi} \int d^4 x \sqrt{-g} (R-2\Lambda)+\int d^4 x \sqrt{-g} L_m \, .
\end{equation}
Assuming the metric  (\ref{metric}) in the gravity part and freezing all other degrees of freedom, we obtain the MSS-action,
\begin{equation}
\label{Sg}
S_{g}= {1\over 2} \int_{t_i}^{t_f} dt \, Na^3 \left( -{1\over N^2}\left({\dot{a}\over a}\right)^2+{1\over a^2}-
  {\lambda} \right) +  {1\over 2} \left[{a^2 \dot{a}\over
    N}\right]_{t_i}^{t_f} \, , \quad \mbox{where} \quad \lambda=
{\Lambda \over 3} \sigma^2 \, .
\end{equation}
In this expression, $a$ and $N$ have been chosen positive, without loss of generality. Furthermore, the boundary terms arise from the integration by
parts of the second derivative of $a$ in the scalar curvature $R$. 

The non-vanishing components of the energy-momentum tensor which are induced by the matter fields (and which are invariant under the
isometries of the RW space-time), can be expressed in terms of
the energy density  $\rho$  and pressure $P$ :  
\begin{equation}
\label{T}
T_{tt}= \rho  \sigma^2 N^2 \, , \quad
T_{ij}= P\sigma^2 a^2 \tilde{g}_{ij}\, ,
\end{equation}
where $d\Omega_3^2=\tilde{g}_{ij}dx^idx^j$ is the line element on
$S^3$. These quantities appear in the equations of motion of $N$ and
$a$ \ie the time-time and space-space Einstein equations : 
 \begin{equation}
 \label{Gtt}
3\left( {\dot{a} \over a} \right)^2+ 3 {N^2 \over a^2} =
12\pi^2\sigma^4 N^2 \rho +3 \lambda N^2  ,
\end{equation}
\begin{equation}
\label{Gii}
-2{\ddot{a} \over a}+2{\dot{a} \over a}{\dot{N} \over N}-\left(
  {\dot{a} \over a} \right)^2-{N^2 \over a^2}=  12\pi^2 \sigma^4 N^2
P- 3\lambda N^2  \, . 
\end{equation}
Since $N$ can be fixed to any arbitrary function by a
reparameterization of $t$ in   
the metric (\ref{metric}), this differential system of two unknown
functions must be degenarate so that we only have to solve any of the
two equations\footnote{This is not true for static solutions.}, e.g. the Friedman one (\ref{Gtt}). Let us consider some cosmological solutions associated to various matter contents and where phase transitions can occur.  


\section{Pure de Sitter space} 
\label{Sitter}

With neither matter nor thermal bath, we have $\rho=P=0$. Choosing $N(t)\equiv 1$, eq. (\ref{Gtt}) is easily solved,
\begin{equation}
\label{deSitter}
a(t)= {\cosh(\sqrt{\lambda}t)\over \sqrt{\lambda}} \, .
\end{equation}
This corresponds to an usual de Sitter evolution where a phase of
contraction from an infinitely past time is followed by an expansion
phase when the radius $a$ has reached its minimum.  

However, a different cosmological scenario can give rise to an
identical de Sitter expansion. Actually,  one can consider the real
time evolution for $t\ge 0$ as the result of a tunnel effect, since an
analytic continuation at $t_i=-i\t_{f}=0$ is allowed. The instanton
solution in question is obtained by substituting $t=-i\t$ in
(\ref{deSitter})\footnote{For any field $f(t)$, we define in general
  its Euclidean counterpart $f_E(\t)\equiv f(-i\t)$ and its derivative
  $\dot{f}_E(\t)= -i\dot f (-i\t)$.},
 \begin{equation}
\label{S4}
a_E(\t)= {|\cos(\sqrt{\lambda}\t)|\over \sqrt{\lambda}} \, ,
\end{equation}
where we find convenient to insist on a conventional positivity of $a_E$
for future purpose. The Euclidean metric takes the form $ds^2_E=
(\s^2/\lambda)(\lambda d\t^2+\cos^2(\sqrt{\lambda}\t) d\Omega_3^2)$, so
that the instanton is a 4-sphere of radius $\s/\sqrt{\lambda}$, if
$\sqrt{\lambda}\t \in [-\pi/2, \pi/2]$.

 Vilenkin, assuming that  the range of the Euclidean time $[\t_i,\t_f]=[-\pi/2 \sqrt{\lambda},0]$, 
considers this instantonic hemisphere as giving rise to a transition
amplitude from nothing (a state of the space-time where it is the empty
set) to an $S^3$ space with de Sitter cosmological evolution
\cite{Vilenkin:1982de, Vilenkin:1983xq}. As we shall see later, this
interpretation is very natural in the deformed de Sitter solution which is induced 
by the thermal bath. The Euclidean action, 
\begin{equation}
\label{SEg}
S_{Eg}= {1\over 2} \int_{\t_{i}}^{\t_{f}} d\t \, N_Ea_E^3\left( -{1\over
    N^2_E}\left( {\dot{a}_E\over a_E}\right)^2-{1\over a^2_E}+ \lambda \right) +  {1\over 2}
\left[{a_E^2 \dot{a}_E\over N_E}\right]_{\t_{i}}^{\t_{f}} \, ,  
\end{equation}
evaluated on shell, has vanishing boundary terms and takes the value
\begin{equation}
\label{Shemisphere}
S_{Eg}= -{1\over 3\lambda}\, ,
\end{equation}
so that the probability transition
\begin{equation}
p\propto e^{2 /(3\lambda)}\, 
\end{equation}
becomes maximal for a vanishing cosmological constant. 
Considering the second hemisphere of the $S^4$-instanton,
$\sqrt{\lambda}\t_{i}=0$ and $\sqrt{\lambda}\t_{f}=\pi/2$, we note
that the previous probability also evaluates the chances for a
contracting de Sitter space to simply vanish instantaneously when it
reaches its minimal radius at $t=0$.

\noindent
{\large \em Another derivation}

\begin{figure}[h!]
\begin{center}
\includegraphics[height=9cm]{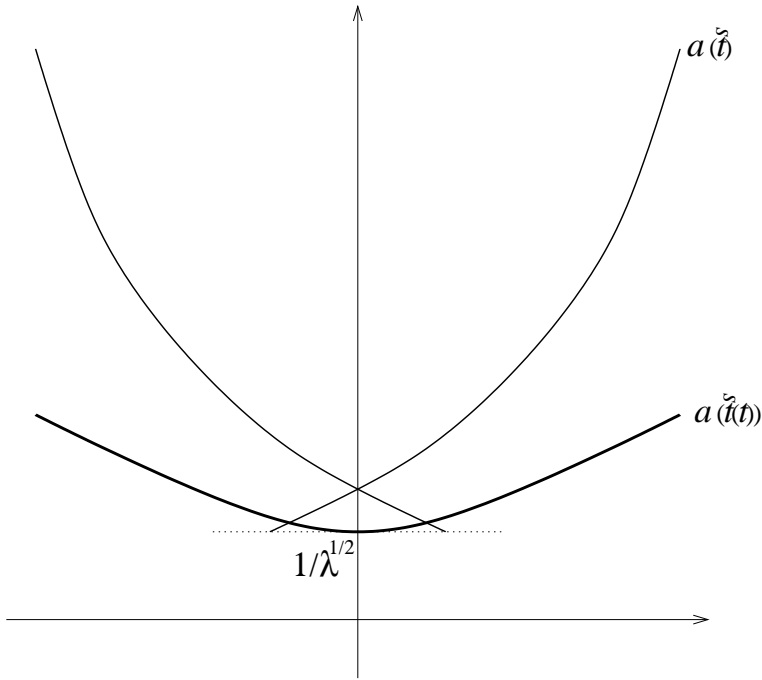}
\caption{\footnotesize \em As a function of $\tt$, the pure de Sitter scale factor $a$ has two
  branches. As a function of $t$, these branches are glued together in a smooth way.}
\label{fig_new_t_de_Sitter}
\end{center}
\end{figure}

\noindent
Before closing this section, we would like to mention an alternative method to solve the
Friedman equation (\ref{Gtt}) that will appear useful to express the general deformed de Sitter solution once the thermal and moduli effects are simultaneously present. Consider 
\be
\label{N}
N^2(\tt)= {1\over 1+{\l^{-1}\k /a^2(\tt)}}\, ,
\ee
and choose  $\k=-1$ so that the $1/a^2$ term of eq. (\ref{Gtt}) is
cancelled out\footnote{As signaled in footnote \ref{fo}, remind
  that the ``dot'' derivative is here with respect to $\tt$, the time
  variable in the gauge (\ref{N}).},  
\be
\left( \ad\over a\right)^2=\l\, . 
\ee
There are two solutions to this equation, whose corresponding metrics are 
\be
ds^2_\pm = \s^2\left( {-d\tt^2\over 1-\l^{-1}e^{\mp2\sqrt\l \tt}}+e^{\pm 2\sqrt\l \tt} d\O_3^2\right)\, ,
\ee
defined for $\tt> -\log(\l)/2\sqrt \l$ and $\tt<\log(\l)/2\sqrt \l$,
respectively. These parameterizations do not admit analytic
continuations to Euclidean time, and actually correspond to ``half
solutions'', as can be seen as follows. Integrating in each case 
\be
\label{newtime}
dt = N(\tt) d\tt\, ,
\ee
one can choose the time variable $t$ (corresponding to the gauge $N\equiv 1$) such that
\be
\label{chvar}
\l e^{\pm2\sqrt \l \tt}=\cosh^2(\sqrt \l t)\, ,
\ee
to glue the two branches in $\tt$ in a smooth way, $t \in \R$ (see
fig. \ref{fig_new_t_de_Sitter}).
In that case, the metrics $ds^2_\pm$ are taking a RW form in
the $N\equiv 1$ gauge (see eq. (\ref{metric})), with $a$ defined in
eq. (\ref{deSitter}). 
%


\section{Thermally deformed de Sitter universe} 

The quantum fluctuations of the previous background
metric have been shown to induce an additional effective
``radiation term" in the action. This implies a back reaction of the
massless gravitationnal degrees of freedom on on the metric \cite{Sarangi:2005cs,  Sarangi:2006eb}. In other words, the solution of the new equations of motion
describes the evolution of a RW closed universe in presence of thermalized
radiation. An explicit
derivation of the effective thermal contribution can be found in
\cite{Sarangi:2005cs, 
  Sarangi:2006eb}, while the consistency of the semi-classical
approach has given a similar result \cite{Brustein:2005yn}. We would like
here to reconsider these effects by using thermodynamical
and precise dynamical constraints to reveal new features. The results
can be applied to any model 
involving massless  bosonic or fermionic modes (beside the
gravitationnal ones), as long
as their vaccum expectation values do not show up in the effective action.

For bosonic (or fermionic) fluctuating states of masses $m_b$ (or
$m_f$) in thermal equilibrium at temperature $T$, the general
expressions of the energy density $\rho_T$ and pressure $P_T$ are 
\begin{equation}
\rho_T = \sum _{\mbox{\scriptsize boson }b} I_\rho^B(m_b) + \sum
_{\mbox{\scriptsize fermion }f}  I_\rho^F(m_f)\; , \quad  P_T = \sum
_{\mbox{\scriptsize boson }b} I_P^B(m_b) + \sum _{\mbox{\scriptsize
    fermion }f}  I_P^F(m_f)\, , 
\end{equation}
where
\begin{equation}
I_\rho^{B(F)}(m_{b(f)})=\int_0^\infty dq {q^2 E(m_{b(f)}) \over
  e^{{1\over T} E(m_{b(f)})}\mp1} \, , \quad I_P^{B(F)}= {1\over 3}
\int_0^\infty dq {q^4 \left/ E(m_{b(f)}) \right. \over e^{{1\over
      T}E(m_{b(f)}) }\mp1  }   
\end{equation}
and $E(m_{b(f)}) = \sqrt{q^2+m_{b(f)}^2}$. If the system is consisting
in $n_{B(F)}$ massless degrees of
freedom, it is easy to see that the state equation 
\begin{equation}
\label{rhoT4}
\rho_T = 3 P_T={\pi^4\over 15}\left(n_B+{7\over 8}n_F\right)  T^4
\end{equation}
is satisfied. These relations  combined with the expression of the
energy-momentum conservation  
\begin{equation}
\label{conservation}
\dot{\rho}_T+3\left({\dot{a}\over a}\right)(\rho_T+P_T)=0
\end{equation}
is then implying that 
\begin{equation}
\label{Ta}
aT=\mbox{constant} 
\end{equation}
and that
\begin{equation}
\label{rho}
\rho_T={\delta^2_T\over 16\pi^2\sigma^4 \l}\, {1\over a^4}\, ,
\end{equation}
where $\delta^2_T/(16\pi^2\s^4\l)$ is a positive integration constant. The factor of $\l$ in its denominator is chosen for later convenience.

To proceed, we solve the Einstein equations in the case where the energy density
and pressure are the thermal ones. This has been done for eq.
(\ref{Gii}) in \cite{Sarangi:2005cs, 
  Sarangi:2006eb, Brustein:2005yn} and \cite{Harrison} in the $N(t)\equiv
1$  gauge. However,
we prefer 
to deal with eq. (\ref{Gtt}) as in \cite{Brout:1989pw} but in a way similar to what we did  at the
end of the last section 
for the pure de Sitter case, since this will be easier to generalize
when we take into account moduli fields. For this purpose, we choose
again $N(\tt)$ with a form 
\be
\label{Nbis}
N^2(\tt)= {1\over 1+{\l^{-1}\k /a^2(\tt)}}\, ,
\ee
and look for $\k$ so that the
$1/a^4$ thermal contribution is cancelled out. This implies that  
\be
\label{2roots}
\k^2+\k+{\d_T^2\over 4}=0\, ,
\ee
a condition that admits real solutions as long as the constraint  $\d^2_T\le 1$ is satisfied. 


\noindent {\large \em The case $\d^2_T <1$}

\noindent The Friedman equation (\ref{Gtt}) becomes
\be
\label{cosh}
\left( {\dot{a} \over a} \right)^2+ { \k+1 \over a^2} = \l \, ,
\end{equation}
showing that the solution of our present problem can be related 
to the one of section \ref{Sitter}, the de Sitter case with no
thermal effect, up to a shift in the $1/a^2$ curvature
term. The hyperbolic cosine solutions of eq. (\ref{cosh}) for the two roots of (\ref{2roots}) are giving rise to
the metrics 
\be
ds^2_\pm = \s^2\left( {-d\tt^2\over 1-{u^{\pm 1}\over \cosh^2(\sqrt \l
      \tt)}}+{1\pm \sqrt{1-\d^2_T}\over2\l} \cosh^2(\sqrt \l
  \tt)d\O_3^2\right)\, , 
\ee 
where
\be
u={1-\sqrt{1-\d^2_T}\over 1+\sqrt{1-\d^2_T}} \in ]0,1]\, ,
\ee
if $\d^2_T>0$. Since they simply correspond to different gauge choices
of functions $N$, we already know that they are equal up to a
redifinition of time. This means that the potential singularity of $ds^2_-$
when the denominator of $N(\tt)$ vanishes should be fake. In any case,
let us make contact with the more intuitive $N\equiv 1$ gauge by
choosing a new time variable $t$ satisfying eq. (\ref{newtime}). This
  will in particular allow us to study in an explicit way various
  analytic continuations.  

\begin{figure}[h!]
\begin{center}
\includegraphics[height=9cm]{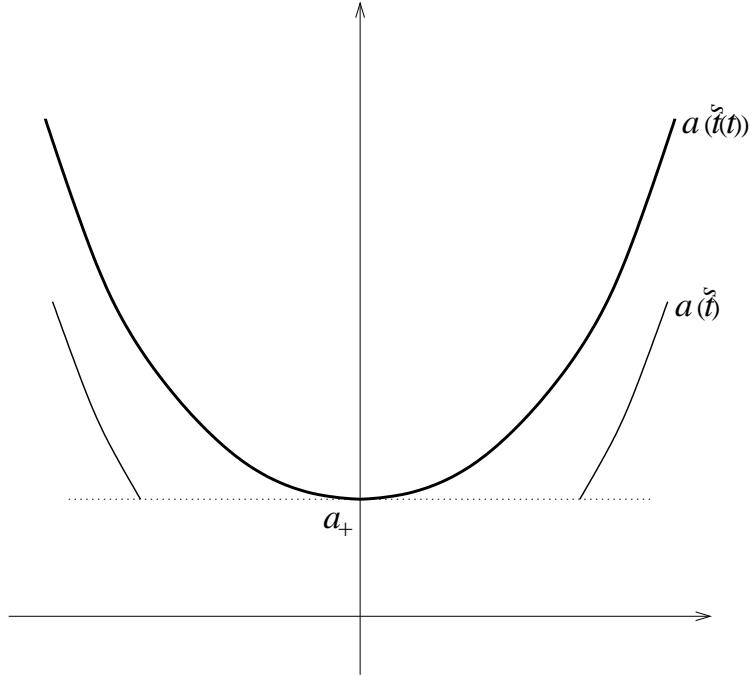}
\caption{\footnotesize  When one is considering the time variable $\tt$
  involved in  $ds^2_-$ in the pure thermal case, the 
  scale factor $a$ has two branches. In terms of $t$, these branches
  are glued together in a smooth way. At the minimum, one has $a=a_+$
  given in eq. (\ref{jump}).}
\label{fig_new_t_therm_de_Sitter}
\end{center}
\end{figure}

For $ds^2_+$, one can fix the integration constant such that
\be
\cosh^2(\sqrt \l \tt) = u+(1-u)\cosh^2(\sqrt \l t)\, ,
\ee
so that the metric is taking the form $ds_+^2=\s^2
\left(-dt^2+a^2\left(\tt(t)\right)d\O_3^2\right)$, with 
\be
\label{exp2}
a=\N \sqrt{\varepsilon+\cosh^2(\sqrt{\lambda}t)} \, , 
\ee
where
\be
\label{Ne}
\N={(1-\delta^2_T)^{1/4}\over \sqrt{\lambda}}\; , \quad
\varepsilon ={1\over 2}\left({1\over
    \sqrt{1-\delta^2_T}}-1\right)\, . 
\end{equation}
For $ds_-^2$, one can have instead
\be
\label{2branches}
\cosh^2(\sqrt \l \tt) = 1+(u^{-1}-1)\cosh^2(\sqrt \l t)\, ,
\ee
which has the effect to glue the two branches $\sqrt
\l \tt>\mbox{arccosh} (u^{-1/2})$ and $\sqrt \l \tt<-\mbox{arccosh}
(u^{-1/2})$  in a smooth way, with $t\in \R$ (see fig. \ref{fig_new_t_therm_de_Sitter}). This time variable, has the property to explicitly show that $ds_-^2\equiv ds_+^2$.  

This solution is a deformation of the de Sitter solution
(\ref{deSitter}) \ie a contraction phase followed by an expansion
one. As before, the latter can also arise by tunnel effect since the
analytic continuation  at $t_i=-i\t_{f}=0$ is still valid. Let us
consider for a moment the relevant instanton solution obtained under
the substitution $t=-i\t$,  
\begin{equation}
\label{aEtempe}
a_E=\N \sqrt{\varepsilon+\cos^2(\sqrt{\lambda}\t)}\, ,
\end{equation}
for  the range of Euclidean time $-\pi/2 \sqrt{\lambda}\le\t\le 0$. 
An important point is that
at $\sqrt{\lambda}\t_{i}=-\pi/2$, $a_E$ is no longer vanishing as in
the birth of the de Sitter space (ref. \cite{Kamenshchik:1995ib} presents another example of this phenomenon). Instead, a second analytic
continuation is allowed,  $\sqrt{\lambda}\t=-\pi/2 +i\sqrt{\lambda}t$,
and another cosmological solution  
\begin{equation}
\label{exp1}
a=\N \sqrt{\varepsilon-\sinh^2(\sqrt{\lambda}t)}\, ,  
\end{equation}
satisfies (\ref{Gtt}) (in the $N\equiv 1$ gauge), for
$-\mbox{arcsinh}\sqrt{\varepsilon} \le \sqrt{\lambda}t \le 0$. Actually, this sinh-solution is connected to the cosh-solution (\ref{exp2}) via the instantonic one (\ref{aEtempe}). The cosmological evolution makes use of all above mention solutions. \\
$\bullet$ The universe starts from a big-bang era at $t  \sim -\l^{-1/2}\mbox{arcsinh}\sqrt{\varepsilon}$, with an evolution following the sinh-solution till $t=0$. \\ 
$\bullet$ At $t=0$, either $i)$ the space starts to contract till a big crunch occurs at
$t=\l^{-1/2}\mbox{arcsinh}\sqrt{\varepsilon}$, or $ii)$ there is a first order phase transition that
changes instantaneously its radius. \\ 
$\bullet$ In case $ii)$, the cosmological evolution for later times, $t \ge 0$, follows the inflationary  cosh-solution.\\
Actually, the very early part of these scenarios (and late time in case $i)$) are not trustable, assuming that the fundamental theory is string like. The degrees of freedom grow exponentially at high temperature, and blow up at the Hagedorn temperature. This effect gives a cut in time before which  the sinh-solution is no more valid (and another cut for late time in case $i)$). A stringy approach is then necessary to cover such ultra hot eras of the universe.

\begin{figure}[h!]
\begin{center}
\includegraphics[height=9cm]{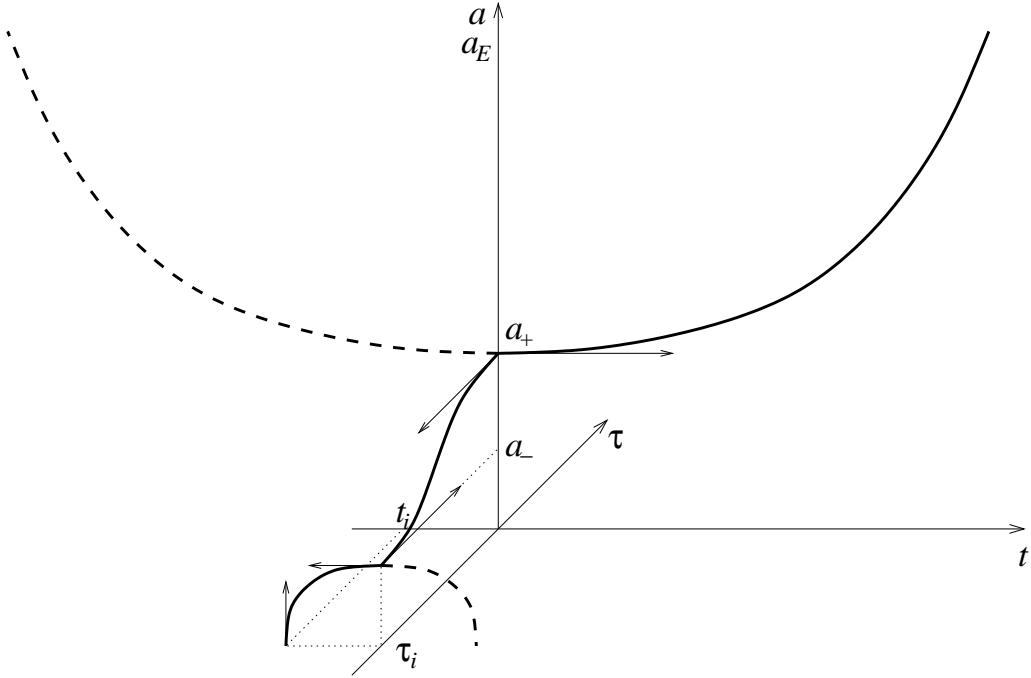}
\caption{\footnotesize \em Cosmological scenario of an expanding RW universe in
  presence of thermal and time-dependent moduli effects below some critical value. A phase of expansion starts from a big bang  and is connected to an
   inflationary deformed de Sitter one by an instanton. This describes a first order transition, where the scale factor  jumps from $a_-$ to $a_+$.}
\label{fig_deformed_cosmology}
\end{center}
\end{figure}

At the transition, the scale factor is jumping,
\be
\label{jump}
a_-=\sqrt{{1-\sqrt{1-\d_T^2}\over 2\l}} \,\,\,\, \longrightarrow  \,\,\,\, a_+=\sqrt{{1+\sqrt{1-\d_T^2}\over 2\l}}\, ,
\ee 
(see fig. \ref{fig_deformed_cosmology}).
The transition probability at $t=0$ (see section \ref{pro}) depends on $\lambda$ and on the deformation parameter $\varepsilon$ (or equivalently on $\delta_T^2$). The maximal value of this probability is for finite $\lambda$ and depends on 
$\varepsilon$. This is fundamentally different from the pure de Sitter case where the maximal  probability is for $\lambda=0$.

\begin{figure}[h!]
\begin{center}
\includegraphics[height=7cm]{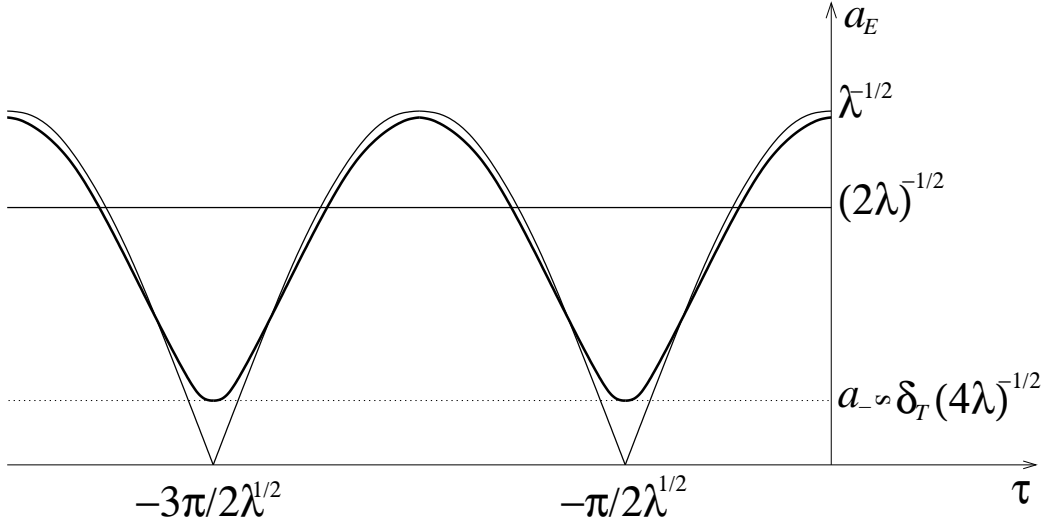}
\caption{\footnotesize \em Thermal instanton (in bold) compared to the
  pure de Sitter $S^4$ one. In the limit where thermal effetcs are
  negligible, $\delta_T \ll 1$, the instantonic jump of $a$ is
  drastic, passing from $a_-\sim \delta_T/2\sqrt{\l}$ to $a_+\sim
  1/\sqrt{\l}$. When thermal effects are getting closer and closer to
  their critical magnitude, 
  $\delta_T =1$, the amplitude of 
  the transition from the big bang branch to the inflationary
  deformed de Sitter one is decreasing. At the very precise value
  $\delta_T =1$, the instanton solution is a constant,
  $a_E(\t)\equiv 1/\sqrt{2\l}$.} 
\label{fig_cos}
\end{center}
\end{figure}

It is interesting to make a comparison between the pure de Sitter case and the 
thermally deformed one. When $\delta^2_T\to 0$, the big
bang occurs at $t_i\sim -\delta_T/2\sqrt{\l}$, while we also have $a_-\sim
\delta_T/2\sqrt{\l}$ at $t=0_-$. This means that Vilenkin's interpretation of the
spontaneous  creation of 
the pure de Sitter space can be seen as the limit when
the big bang sinh-cosmological branch has shrunk to nothing (see fig. \ref{fig_cos}). 
Actually, the instantonic jump $a_-\rightarrow a_+$ becomes more drastic as the
thermal effects become more negligible, $\delta_T\ll 1$, since $a_+\sim
1/\sqrt \l$ in this limit. On the contrary, the transition from the big bang branch to the cosh-inflationary
deformed de Sitter one is smoother and smoother as we approach the
critical magnitude of the thermal effects, $\delta_T\to 1$. Actually
the amplitude of the instanton solution decreases as we approach this
value of the parameter. When $\delta_T=1$, the instanton solution
is constant, implying that its analytic continuations are static,
$a(t)\equiv 1/\sqrt{2\l}$.    

\begin{figure}[h!]
\begin{center}
\includegraphics[height=12cm]{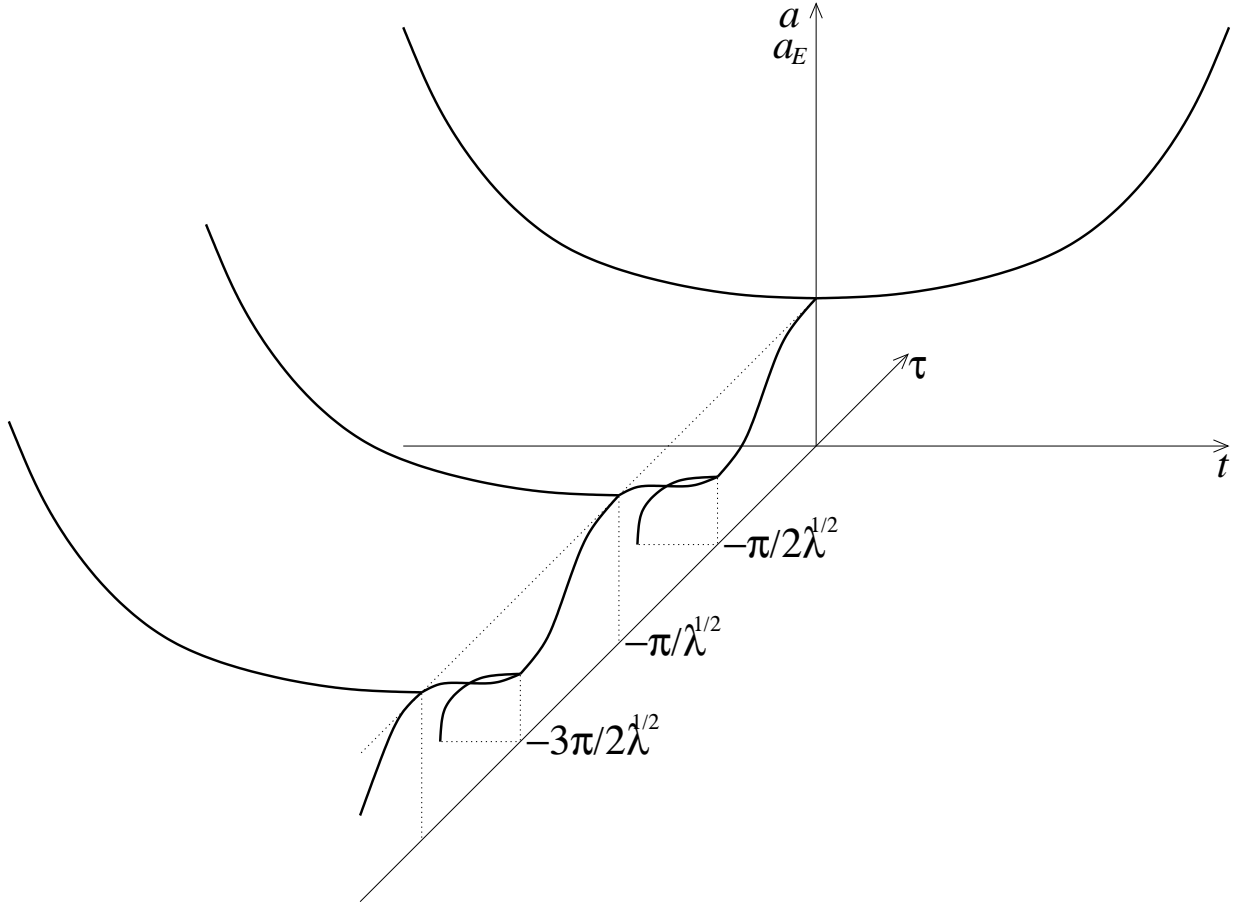}
\caption{\footnotesize \em The inflationary phase of the deformed de
  Sitter space for $t\ge 0$ can arise from any contracting phase or
  big bang phase for $t\le0$, after a multi-instantonic transition.}
\label{fig_multi_instantons}
\end{center}
\end{figure}

An estimate of the transition probability from the big bang branch to
the deformed de Sitter one is $p\propto e^{-2S_{E\mbox{\tiny
      \em{eff}}}}$, where  
\begin{equation}
\label{SEtemp}
S_{E\mbox{\tiny \em{eff}}}= {1\over 2} \int_{\t_{i}}^{\t_{f}} d\t \, N_E a_E^3
\left( -{1\over N^2_E}\left(\dot{a}_E\over a_E\right)^2-{1\over a_E^2}+ \lambda +
  {\delta^2_T \over 4\l a^4_E}\right)  +  {1\over 2}
\left[{a_E^2 \dot{a}_E\over N_E}\right]_{\t_{i}}^{\t_{f}}
\end{equation}
is the Euclidean effective action that is giving rise to the desired
equations of motion. In this expression, $S_{E\mbox{\tiny \em{eff}}}$
is evaluated with the solution (\ref{aEtempe}) and has vanishing boundary terms for $-\pi/2\le
\sqrt{\lambda}\t\le 0$, (see eq. (\ref{acT})). However, one could sum in principle the
instanton contributions for the series of ranges $-(2n+1)\pi/2\le
\sqrt{\lambda}\t\le 0$, $n\in \intN$. The total probability would take
the form $p_{\mbox{\tiny \em tot}}=\sum_{n\ge 0}c_{2n+1}
e^{-(2n+1)2S_{E\mbox{\tiny \em{eff}}}}$, where $c_{2n+1}$ are
positive constants. Similarly, the probability to pass from the
contracting phase of the deformed de Sitter space to the expanding
one would take the form $p'_{\mbox{\tiny \em tot}}=\sum_{n\ge
  0}c_{2n} e^{-4nS_{E\mbox{\tiny \em{eff}}}}$, for positive
constants $c_{2n}$, (see fig. \ref{fig_multi_instantons} and \cite{Barvinsky:2006uh}).  
%


\noindent {\large \em Introducing the temperature at the transition}

\noindent It is useful to express the deformation parameter $\delta^2_T$ (or $\N$ and $\varepsilon$ in eq. (\ref{Ne})) in terms of more intuitive quantities, such as  temperatures. First, observe that before and after the transition, the quantity  $aT$ remains  constant (see eq. (\ref{Ta})). Since these evolutions are related by a double analytic continuation, the constants must be equal. This allows us to
express $aT$ in terms of either the temperatures $T_+$ at $t=0_+$ or
$T_-$ at $t=0_-$ : 
\begin{equation}
\label{Tback}
T={T_\pm\over a} {1\over \sqrt{2\lambda}}
\sqrt{1\pm\sqrt{1-\delta^2_T}}\, . 
\end{equation}
Plugging either of these relations in the expression of $\rho_T$ in
eq. (\ref{rhoT4}) and comparing the result with eq. (\ref{rho})
imposes  a consistency condition to be satisfied by
$\delta^2_T$. Its solution involves a mean temperature $T_m$
satisfying $T_+\le T_m\le T_-$, 
\be
\label{Tm}
\delta^2_T={4\over \left( T^2_m/T^2_\pm+T^2_\pm/
    T^2_m\right)^{2}}\quad \mbox{where} \quad T_m= 
\left({15\over 4\pi^6} {\l\over (n_B+{7\over 8}n_F) }\right)^{1/4}
{1\over \s}\, . 
\ee
This implies that $T_+T_-=T_m^2$ and that  $\delta^2_T$ is invariant
under the duality  transformation 
\be
\label{sim}
{T_+\over T_m}\leftrightarrow {T_m\over T_+}\equiv {T_-\over T_m}\, .
\ee
Actually, this operation exchanges the scale factor of the thermally deformed
de Sitter evolution,
\be
a={1\over\sqrt{2\l}}\sqrt{1+{T_m^4-T_+^4\over T_m^4+T_+^4}\cosh 
  (2\sqrt\l t)}\; , \quad t > 0\, ,
\ee
with the big bang / big crunch one, 
\be
a={1\over\sqrt{2\l}}\sqrt{1+{T_m^4-T_-^4\over T_m^4+T_-^4}\cosh 
  (2\sqrt\l t)}\; , \quad t < 0. 
\ee
In some sense, the instantonic
transition between them is  ``inversing'' the
temperature, since $T(t)\ge T_-$ for $t\le 0$, while
$T(t)\le T_+=T^2_m/T_-$ for $t\ge 0$. If one wishes, one can also
rewrite the constants $\N$ and $\ve$ of eq. (\ref{Ne}) in
terms of either $T_+$ or $T_-$, 
\be
\N={1\over \sqrt{\l}}\sqrt{{1-(T_\pm/T_m)^{\pm 4}\over
    1+(T_\pm/T_m)^{\pm 4} }}\; , \quad \ve= {1 \over (T_\pm/T_m)^{\mp 4}-1}\, ,
\ee
and note that the function $N^2$ appearing in eq. (\ref{Nbis}) satisfies $N^2(\tt)= 1/(1-T^2(\tt)/T_\mp^2)$ for $ds^2_\pm$, respectively.  


\noindent {\large \em Focussing on the $\l$-dependance of $\delta_T^2$}

\noindent We would like here to consider the dependance on $\l$ of $\delta_T^2$. This is relevant, for instance, if one wants to minimize the Euclidean action with respect to $\l$ (keeping fixed the number of degrees of freedom and temperatures $T_\pm$). Doing so was considered as a selection principle for the cosmological constant in \cite{Firouzjahi:2004mx, Sarangi:2005cs, Sarangi:2006eb, Brustein:2005yn, BouhmadiLopez:2006pf}. The reason for this was based on the fact that  this action (see eq. (\ref{acT})) is controlling the probability transition between the two cosmological evolutions. However, there are two hypothesis for this argument to be considered. The first one is that the viable universes are the deformed de Sitter ones, \ie that the big bang / big crunch evolution is too hot or too short in time. However, this condition implies that $\delta_T^2$ is small enough, which in turn can be translated into a condition on $\l$ itself (see eq. (\ref{Tm})). The second is that  the probability to start the $\l$-dependent big bang evolution is almost constant as a function of $\l$. 

Eq. (\ref{Tm}) can be rewritten as 
\be
\label{nu}
\delta^2_T = {\nu_+\over \l}{4\over \left( 1+{\nu_+ / \l}\right)^2}= {\l\over \nu_-}{4\over \left( 1+{\l / \nu_-}\right)^2}\, ,
\ee
where
\be
\nu_\pm = {4\pi^6\over 15}\left(n_B+{7\over 8}n_F\right)T_\pm^4\s^4\, ,
\ee
and reproduces the result of \cite{BouhmadiLopez:2006pf}.
Actually, $\l$ is the geometric mean of $\nu_+$ and $\nu_-$,
\be
\label{cond}
\left({T_+ \over T_m}\right)^4={\nu_+\over \l}={\l\over \nu_-}=
\left({T_m \over T_-}\right)^4\le 1\, ,
\ee
so that eq. (\ref{nu}) is relevant to derive an expansion of $\delta_T^2$
for small $\nu_+/\l$, whose leading term reproduces the result of  
\cite{Sarangi:2005cs, Sarangi:2006eb, Brustein:2005yn}.  Eq. (\ref{nu}) goes beyond the small
$\nu_+/\l$ approximation by taking consistently into account  the back reaction term $\delta_T^2$ that appears in the r.h.s. of eq. (\ref{Tback}). 

A conceptual difference of our approach compared to that of \cite{Sarangi:2005cs, Sarangi:2006eb, Brustein:2005yn} is that we insist on the difference between the two temperatures $T_{\pm}$, where  $T_- \ge T_+ $. Thus, the only stringy constraint to not reach the Hagedorn temperature during the cosmological evolution is $T_-\ll T_H$. In that case, the temperature $T(t)$ remains small, $T(t)\ll T_H$, for all $t> t_H$ (and $t<-t_H$ in the big crunch branch), where  $t_H$ is the Hagedorn time. 


\noindent {\large \em The case $\d^2_T >1$}

\begin{figure}[h!]
\begin{center}
\includegraphics[height=9cm]{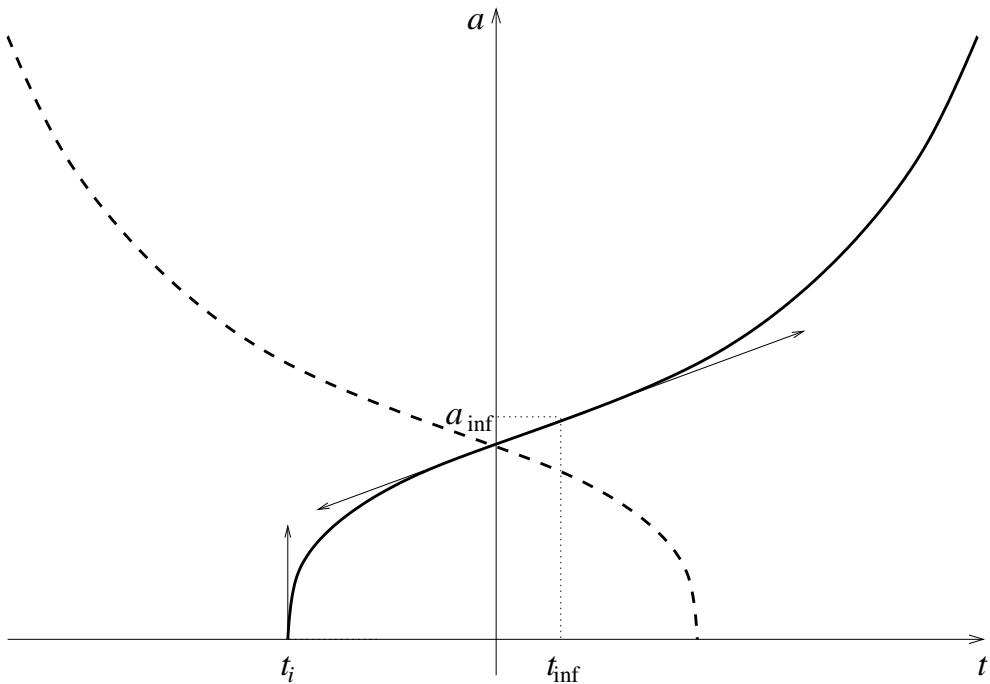}
\caption{\footnotesize \em Cosmological scenario of an expanding RW universe in
  presence of thermal and 
  non trivial time-dependent moduli effects above some critical value. A phase of expansion starts from a big bang  and evolves smoothly toward an
   inflationary one. This describes a second order transition.}
\label{fig_deformed_cosmology_2}
\end{center}
\end{figure}

\noindent We cannot use any more a function $N$ of the form (\ref{Nbis}) to map the Friedman equation of the present case to the pure de Sitter one. We thus actually solve eq. (\ref{Gtt}) in a more straightforward way. In the gauge $N\equiv 1$, one can immediately express a monotically increasing time $t$ as a function of $a$ :
\be
t(a)= \int_0^a{u\, du\over \sqrt{\l u^4-u^2+\delta_T^2 /4\l}}+t_i\equiv {1\over 2\sqrt{\l}}\, \mbox{arctanh}\left({2\l a^2-1\over 2\sqrt{\l}\sqrt{\l a^4-a^2+\delta_T^2/4\l}}\right)\, , \quad a\ge 0\, ,
\ee
where the argument of the square root is never vanishing and we have chosen the integration constant $t_i=-(4\l)^{-1/2}\mbox{arctanh}\, \delta_T^{-1}$. Inverting the function $t(a)$, one finds
\be
\label{order2}
a(t)={1\over \sqrt{2\l}}\, \sqrt{1+\sqrt{\delta_T^2-1}\, \sinh(2\sqrt{\l}\, t)}\; , \quad t\ge t_i\equiv -{1\over 2\sqrt{\l}}\, \mbox{arcsinh}\, {1\over \sqrt{\delta_T^2-1}}\, .
\ee
The cosmological evolution described by this solution starts with a big bang at $t=t_i$, while for large positive time, it is inflationary. It thus looks similar to the case $\delta_T^2<1$, both at the beginning of time and for large $t$, when the first order transition has occurred. However, in the present case, the evolution from the small to very large scale factors is smooth, (see fig. \ref{fig_deformed_cosmology_2}). 
Actually, the solution (\ref{order2}) has an inflection point  arising at $t=t_{\mbox{\tiny inf}}$, where $a(t_{\mbox{\tiny inf}})=a_{\mbox{\tiny inf}}$,
\be
t_{\mbox{\tiny inf}}={1\over 2\sqrt{\l}}\, \mbox{arcsinh}\sqrt{{\delta_T-1\over \delta_T+1}}\; , \quad a_{\mbox{\tiny inf}}=\sqrt{{\delta_T\over 2\l}}\, ,
\ee
associated to a second order phase transition.   As in the case $\delta_T^2<1$, the solution should be trusted as soon as the temperature is below $T_H$.

For completeness, we signal that another solution to eq. (\ref{Gtt}) is found under the time reversal $t\to -t$ and is thus monotically decreasing. A de Sitter like universe is contracting since an infinitely past time and evolves in a smooth way around $t=t_{\mbox{\tiny inf}}$ toward a branch that ends with a big crunch.


\noindent {\large \em The case $\d^2_T =1$}

\noindent Taking the limit $\delta_T^2\to1$ in the solutions found in the cases $\delta_T^2 <1$ and $\delta_T^2> 1$ reaches the same static solution,
\be
a(t)\equiv a_0\quad \mbox{where} \quad a_0={1\over \sqrt{2\l}}\, .
\ee
It corresponds to a universe with an $S^3$-space of constant radius. Actually, the $\delta_T^2\to1_-$ limit amounts to send the big bang to an infinitely past time  (see eqs. (\ref{Ne}), (\ref{exp1})), while keeping the instant the transition occurs at $t=0$. Similarly, the $\delta_T^2\to1_+$ limit sends $t_i\to -\infty$ (see eq. (\ref{order2})), while keeping $t_{\mbox{\tiny inf}}$ finite. In both cases, the solutions are thus ``stretched" and converge toward a constant. In the two cases, it is however possible to redefine the big bang time at a conventional $t=0$ by a $\delta_T^2$-dependent shift in the definition of time. Taking only then the limit   $\delta_T^2\to 1$ reaches solutions where the dynamical part of the solutions are not sent to infinity any more. 

\begin{figure}[h!]
\begin{center}
\includegraphics[height=9cm]{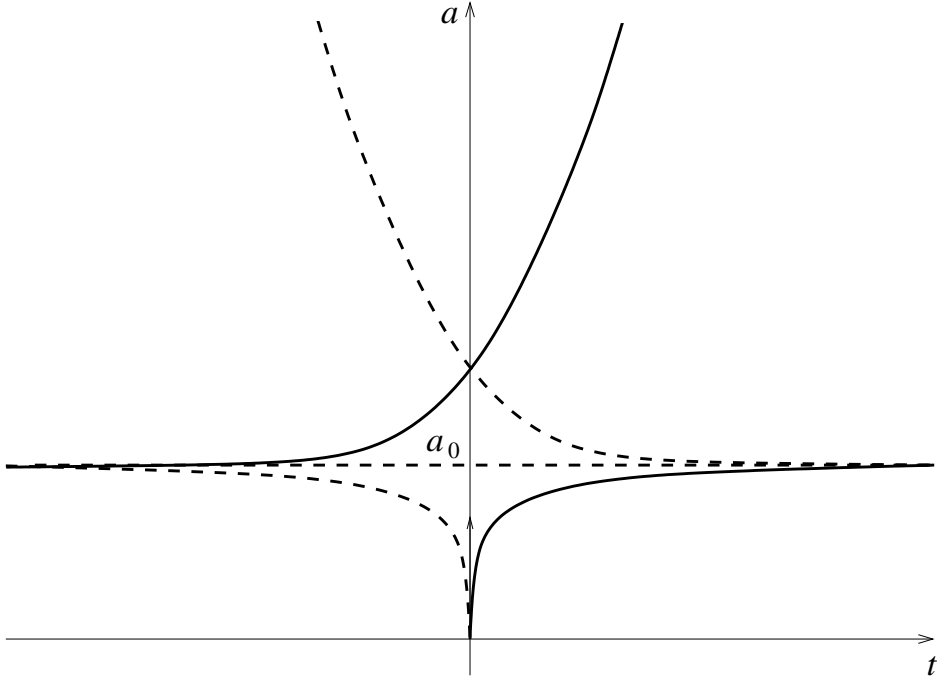}
\caption{\footnotesize \em  Cosmological scenario of an expanding RW universe in
  presence of thermal and non trivial time-dependent moduli effects at a critical value. A phase of expansion starts from a big bang  and converges quickly toward a static universe. After a very long time, the universe can either die in a big crunch or enter into an inflationary phase.}
\label{fig_deformed_cosmology_3}
\end{center}
\end{figure}

To find these new solutions, one can write the Friedman equation (\ref{Gtt}) in the form
\be
\left( a\dot a\right)^2=\l (a^2-a_0)^2\, ,
\ee
and derive two monotically increasing  solutions for the time as a function of $a$,
\be
t(a)={1\over \sqrt{\l}}\int_0^a{u\, du\over a_0^2-u^2}\; , \quad 0\le a<a_0\, ,
\ee
and
\be
t(a)={1\over \sqrt{\l}}\int_{2a_0}^a{u\, du\over u^2-a_0^2}\; , \quad a>a_0\, .
\ee
Integrating explicitly these expressions and inverting the functions $t(a)$ reaches then the two cosmological evolutions :
\be
\label{<}
a(t)=a_0 \sqrt{1-e^{-2\sqrt{\l}\, t}}\; , \quad t\ge 0\, ,
\ee
and
\be
\label{>}
a(t)=a_0 \sqrt{1+e^{2\sqrt{\l}\, t}}\; , \quad t\in \R\, .
\ee
The first one starts with a big bang arising at $t=0$, converges quickly toward the static solution, (and should be trusted as soon as $T(t)<T_H$). The second is a deviation from the static solution toward the inflationary phase. Two other monotically decreasing solutions are found under the time reversal $t\to -t$ in eqs. (\ref{<}) and (\ref{>}). The four solutions are drawn on fig. \ref{fig_deformed_cosmology_3}.
%


\section{De Sitter deformation by moduli fields}
\label{ax}

In this section, we generalize  the thermally deformed de Sitter
solution by including the 
effect of the non trivial time dependence of some moduli fields. 
To be specific, our choice of matter action to be added to (\ref{Sg}) is
\begin{equation}
S_m={M_p^2\over 16\pi}\int d^4x \sqrt{-g} \left(
 -g^{\mu\nu}\partial_\mu\chi_i\partial_\nu\chi_i
 \right) \, ,
\end{equation}
where the $\chi_i$'s  are the moduli fields with neither potential, nor
interactions with 
the thermal system considered before. Assuming the RW metric
(\ref{metric}) and the
 presence of the thermal corrections discussed before,  the modified  MSS  action becomes
\begin{equation}
\label{Sgene}
S_{\mbox{\tiny \em{eff}}}= {1\over 2}  \int_{t_i}^{t_f} dt \, Na^3\left( -{1 \over
    N^2}\left(\dot{a}\over a\right)^2+{1\over a^2}- {\lambda}  - {\delta'^2_T \over 4\l a^4}+
  {1\over6N^2}\dot{\chi}_i ^2\right)  +  {1\over 2} \left[{a^2 \dot{a}\over
    N}\right]_{t_i}^{t_f} \, , 
\end{equation}
where $\delta'^2_T$ is a new positive constant. The energy-momentum
tensor can be  expressed in terms of  
\begin{equation}
\label{rhoex}
\rho={1\over 12\pi^2 \sigma^4 N^2}\left( {3\delta'^2_T \over 4\l }\, {N^2\over a^4}+
  {1\over 2}\dot{\chi}_i^2\right)\, , 
\end{equation}
\begin{equation}
\label{Pex}
P={1\over 12\pi^2 \sigma^4 N^2}\left({\delta'^2_T \over 4\l}\, {N^2\over a^4}+
  {1\over 2}\dot{\chi}_i^2\right)\, . 
\end{equation}
The equation of motion of $\chi_i$ is trivially solved,
\begin{equation}
\dot{\chi}_i={2\sqrt{2}\delta_{\chi_i} \over 3\l} \, {N\over a^3} \, ,
\end{equation}
where $2\sqrt{2}\delta_{\chi_i}/3\l$ is an integration constant, whose form is chosen for later convenience. Defining
\be
\dc=\sum_i{\delta^2_{\chi_i}}\, ,
\ee
the moduli contributions of $\rho$ and $P$ become
\begin{equation}
\label{rhoax}
\rho_\chi = P_\chi ={\delta^2_\chi\over 27\pi^2 \sigma^4 \l^2}\, {1\over  a^6} \, .
\end{equation}

\begin{figure}[h!]
\begin{center}
\includegraphics[height=7cm]{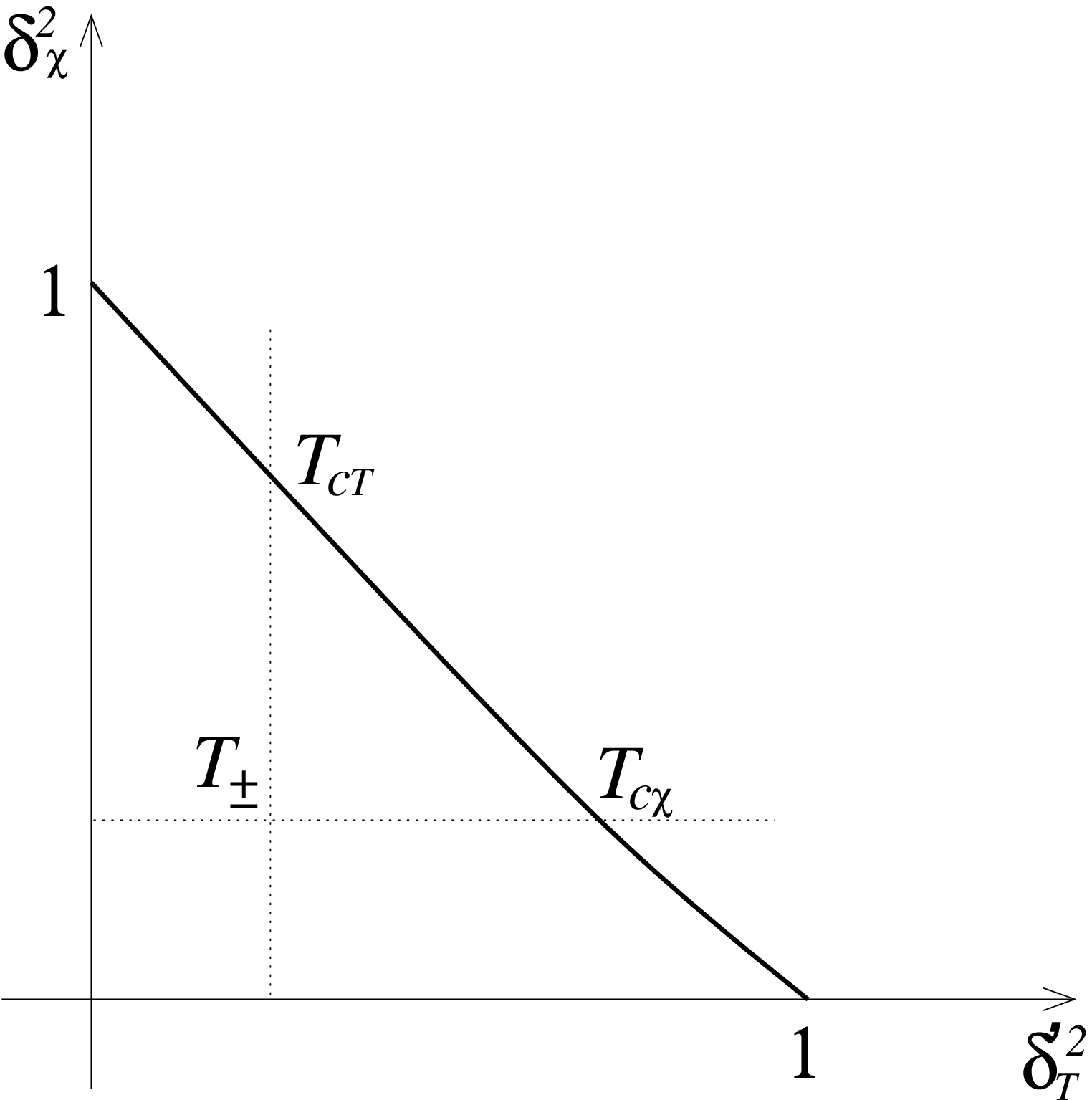}
\caption{\footnotesize \em Phase diagram in the $(\dt,\dc)$-plane. In the domain $\dc\le h(\dt)$ defined  in eq. (\ref{domain}), a first order phase transition between two cosmological branches can occur.  Outside the domain, there is a solution describing a second order phase transition. On the critical curve $\dc= h(\dt)$, there are cosmological evolutions asymtotic to a static solution. Furthermore,  one can introduce the temperatures $T_+<T_-$ in the interior of the domain to replace the two parameters $\dc$ and $\dt$. On the boundary curve, these temperatures converge and their common value can be considered as a function $T_{cT}(\dt)$ or $T_{c\c}(\dc)$.} 
\label{domain_fig}
\end{center}
\end{figure}

To solve the Friedman equation (\ref{Gtt}) in presence of the extra $1/a^6$-term induced by the 
motion of the $\chi_i$ fields, we follow the method implemented in the previous sections.
We look for a real constant $\k'$ in the definition of $N(\tt)$, 
\be
\label{N'}
N^2(\tt)= {1\over 1+{\l^{-1}\k' / a^2(\tt)}}\, ,
\ee  
so that the extra $1/a^6$ contribution is 
cancelled. This can be achieved by choosing $\k'$ such that  
\be 
\label{d3}
\P(\k')\equiv \k'^3+\k'^2+{\d'^2_T\over 4}\k'-{4\over 27}\dc=0 \, .
\ee
This polynomial always has a real positive root (see the appendix). Using it,  eq. (\ref{Gtt}) is taking the form 
\be
\label{defcosh}
\left( {\dot{a} \over a} \right)^2+ {1+\k' \over a^2} = \l +{\k'^2+\k'+\d'^2_T/4\over \l a^4}.
\ee
The above equation has the same form with that of the $1/a^4$-thermally deformed de Sitter one
discussed previously, up to a shift in both coefficients of the $1/a^2$-curvature term 
and the $1/a^4$-thermal term. Using a simple rescaling of the radius $a$ (by a factor $\sqrt{1+\k'}$), one can immediately deduce the solutions from the previous section in terms of the time variable $\tt$. This implies in particular that the analogue of the order parameter $\delta_T^2$ of the pure thermal case is now the positive quantity
\be 
\label{Delta}
\D= {4\k'^2+4\k'+\d'^2_T\over (1+\k')^2}={16\over 27}{\d^2_\c\over  \k'
  (1+\k')^2}\, ,
\ee
where these two definitions of $\D$ are equivalent as a consequence of eq. (\ref{d3}). 
In the appendix, we shown that the condition $\D < 1$ is
equivalent to the statement that the parameters $\dt$ and $\d^2_\chi$
satisfy the inequalities, 
\be
\label{domain}
 \dt < 1 \quad \mbox{and} \quad \d_\c^2 < h(\dt)\, ,
\ee
where
\be
\label{h}
h(\dt)= {1\over 2}\left( \sqrt{1-{3\over 4}\dt}+1\right)^2\left(
  \sqrt{1-{3\over 4}\dt} - {1\over 2} \right)\, . 
\ee
This defines a domain in the  $(\dc,\dt)$-plane shown in
fig. \ref{domain_fig}. At the origin of this domain, one has $\D=0$, while $\D=1$ on its boundary. Outside, this region, on has $\D>1$.
%


\noindent {\large \em The case $\D< 1$}

\noindent Eq. (\ref{defcosh}) admits the cosmological evolution
\begin{equation}
\label{sol}
a(\tt)= \N'\sqrt{\varepsilon' + \cosh^2(\sqrt \l \tt)}\, ,
\end{equation}
where
\be
\label{epsneg}
 \N' = \sqrt{1+\k'\over \l}(1-\D)^{1/4}\quad \mbox{and}\quad
 \ve'={1\over 2}\left({1\over \sqrt{1-\Delta}}-1\right)\, . 
\ee
To proceed, we switch to the more intuitive $N\equiv 1$ gauge.
This requests a redefinition of the time variable dictated  by  eq. (\ref{newtime}) : 
\be
\label{t1}
t=\int_0^{\tt} dv \sqrt{\cosh^2(\sqrt \l v)+\ve'\over \cosh^2(\sqrt \l
  v)+\ve'+\ve^{\prime\prime}} \quad \mbox{where}\quad
\ve^{\prime\prime}={\k' \over (1+\k')\sqrt{1-\Delta}}\, . 
\ee
An explicit form of $t$ as a function of $\tt$ can actually be determined,
however we find more convenient to work with the above integral representation. 
The scale factor $a(\tt(t))$ takes a simple form in terms of the function $\tt (t)$ found by inverting the definition (\ref{t1}), 
\be
\label{sol1}
a(\tt(t)) = \N'\sqrt{\varepsilon'+\cosh^2(\sqrt{\lambda}\,\tt(t))}\, .
\ee
Some comments are in order :\\
$\bullet$ The slope  $d\tt/dt$ is positive everywhere, and  for large $\tt$ one has $t\sim \tt$. This means that the moduli deformation of the thermal de Sitter solution 
appears as a mild redefinition of time and normalization factor. \\
$\bullet$ An Euclidean solution is defined via the analytic continuation, $\tt=-i\taut$ \ie  $t=-i\t$, 
\be 
\label{aE}
a_E(\taut(\t)) =
\N'\sqrt{\varepsilon'+\cos^2(\sqrt{\lambda}\, \taut(\t))}\quad
\mbox{with}\quad \t=-\int^0_{\taut} dv \sqrt{\cos^2(\sqrt \l
  v)+\ve'\over \cos^2(\sqrt \l v)+\ve'+\ve^{\prime\prime}}.
\ee
The range of Euclidean time is 
\be
-\pi/2\le \sqrt \l  \taut \le 0 \quad \mbox{\ie} \quad \t_i \le  \t\le 0\, , 
\ee
where $\t_i$ is given by
\be
\t_i=
-\int^0_{-{\pi\over 2\sqrt\l}} dv \sqrt{\cos^2(\sqrt \l v)+\ve'\over
  \cos^2(\sqrt \l v)+\ve'+\ve^{\prime\prime}}.
\ee
The boundary terms in eq. (\ref{SEg}) vanish.\\
$\bullet$ The Euclidean solution admits, (as in the pure thermal case), two analytic continuations, 
one for each boundary. The one associated to  $\t_i$,  $\sqrt{\lambda}\taut=-\pi/2
+i\sqrt{\lambda}\tt\; $ \ie $\sqrt{\lambda}\t=\sqrt{\lambda}\t_i
+i\sqrt{\lambda}t$, 
gives  rise to a sinh-type cosmological solution,
\begin{equation}
\label{exp3}
a(\tt(t))=\N'\sqrt{\varepsilon'-\sinh^2(\sqrt{\lambda}\, \tt(t))}\quad
\mbox{with}\quad   t=-\int^0_{\tt} dv \sqrt{\ve'-\sinh^2(\sqrt \l
  v)\over \ve'+\ve^{\prime\prime}-\sinh^2(\sqrt \l v)}\, . 
\end{equation}
The range of time is 
\be 
\label{ti}
-{1\over \sqrt{\lambda}}\mbox{arcsinh}\sqrt{\varepsilon'} \le\tt \le 0\quad
\mbox{\ie}\quad t_i=-\int^0_{-{\mbox{\scriptsize
      arcsinh}\sqrt{\varepsilon'}\over \sqrt \l}} dv
\sqrt{\ve'-\sinh^2(\sqrt \l v)\over
  \ve'+\ve^{\prime\prime}-\sinh^2(\sqrt \l v)}\le t\le 0\, . 
\ee

The full scenario
is thus similar to the one presented in the previous section. There is
a big bang at 
$t=t_i$ and the universe expands till $t=0$, before either contracting
or experimenting a first order transition to an inflationary phase. At the
transition, the scale factor is jumping, 
\be
a_-=\sqrt{(1+\k') {1-\sqrt{1-\D}\over 2\l}}  \,\,\,\, \longrightarrow \,\,\,\,
a_+=\sqrt{(1+\k') {1+\sqrt{1-\D}\over 2\l}} .
\ee
Thus, the present moduli deformation  is generalizing the pure thermal one 
discussed previously, (see fig. \ref{fig_deformed_cosmology}). 
  

\noindent {\large \em The case $\D> 1$}

\noindent Eq. (\ref{defcosh}) admits the $\tt$-dependent solution
\be
\label{SOL2}
a(\tt)=\sqrt{1+\k' \over 2\l} \, \sqrt{1+\sqrt{\D-1}\, \sinh(2\sqrt{\l}\, \tt)}\; , \quad \tt\ge \tt_i\equiv -{1\over 2\sqrt{\l}}\, \mbox{arcsinh}{1\over \sqrt{\D-1}}\, ,
\ee
we want to express in terms of a time variable associated to the natural $N\equiv 1$ gauge. Integrating eq. (\ref{newtime}), one can define  $t$ as a function of $\tt$ by
\be
t=\int_0^{\tt} dv \sqrt{\sqrt{\D-1}\, \sinh(2\sqrt \l
  v)+1\over \sqrt{\D-1}\, \sinh(\sqrt 2\l v)+1+{2\k'\over 1+\k'}}\, ,
 \ee
where
 \be
 \label{tbb}
t\ge t_i\equiv -\int^0_{\tt_i} dv \sqrt{\sqrt{\D-1}\, \sinh(2\sqrt \l
  v)+1\over \sqrt{\D-1}\, \sinh(\sqrt 2\l v)+1+{2\k'\over 1+\k'}}\, ,
\ee
and rewrite
\be
a(\tt(t))=\sqrt{1+\k \over 2\l} \, \sqrt{1+\sqrt{\D-1}\, \sinh(2\sqrt{\l}\, \tt(t))}\, .
\ee
As in the previous case, the slope $d\tt/dt$ is always positive and $t\sim \tt$ for large $\tt$. The solution is thus a mild deformation of the pure thermal case. It starts with a big bang at $t=t_i$ and is monotically increasing. For large time, the behavior is inflationary. The evolution can again be interpreted as a second order phase transition since there is a unique inflection point, (see fig. \ref{fig_deformed_cosmology_2}). To show this, one can express $dt/da$ as a function of $a$ from the Friedman equation (in the $N\equiv 1$ gauge) and take a derivative. One finds that the inflection point arises at the radius $a_{\mbox{\tiny inf}}$, where
\be
\l a_{\mbox{\tiny inf}}^2 >0 \quad \mbox{satisfies} \quad x^3-{\dt\over 4}x-{8\over 27}\dc=0\, .
\ee   
Finally, a monotically decreasing cosmological solution is found under the time reversal $t\to -t$.
  

\noindent {\large \em The case $\D= 1$}

\noindent The static solution obtained in the limit $\D\to 1$ in the previous cases is
\be
\label{Static}
a(t)\equiv a_0 \quad \mbox{where}\quad a_0=\sqrt{{1+\k' \over 2\l}}\, .
\ee
Beside this constant radius universe, we again have the two $\tt$-dependent evolutions
\be
\label{<&>}
a(\tt)=a_0 \sqrt{1\mp e^{\mp2\sqrt{\l}\, \tt}}\, .
\ee
To consider them in the $N\equiv 1$ gauge, we integrate eq. (\ref{newtime}) in each case and find  
\be
\label{inf}
a(\tt(t))=a_0 \sqrt{1- e^{-2\sqrt{\l}\, \tt(t)}}\quad \mbox{where}  \quad t=\int_0^{\tt}dv \sqrt{{1-e^{-2\sqrt{\l}v}\over 1+{2\k'\over 1+\k'}-e^{-2\sqrt{\l}v}}}\, ,
\ee
for $\tt\ge0$ \ie $t\ge0$, and
\be
\label{sup}
a(\tt(t))=a_0 \sqrt{1+ e^{2\sqrt{\l}\, \tt(t)}}\quad \mbox{where} \quad t=\int_0^{\tt}dv \sqrt{{1+e^{2\sqrt{\l}v}\over 1+{2\k'\over 1+\k'}+e^{2\sqrt{\l}v}}}\, ,
\ee
for arbitrary $\tt$ and $t$. As in the cases $\D\neq 1$, one has $d\tt/dt>0$ and this redefinition of time does not change the qualitative behavior of the solutions, (see fig. \ref{fig_deformed_cosmology_3}). They are monotically increasing as in the pure thermal case, while two other solutions obtained under $t\to-t$ are decreasing.  


\subsection{Other parameterizations of the solutions when $\D\le 1$}

\noindent {\large \em The temperatures at the 1$^{st}$ order transition}

\begin{figure}[h!]
\begin{center}
\includegraphics[height=8cm]{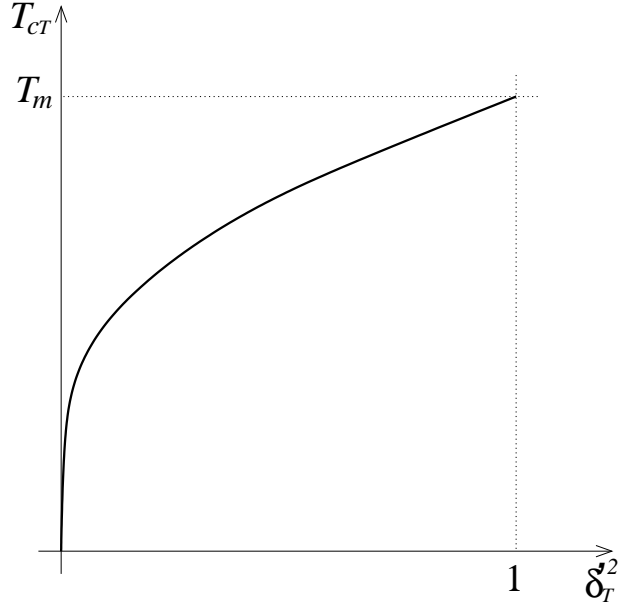}
\caption{\footnotesize \em Critical temperature as a function of $\dt$. It is defined on the curve $\dc=h(\dt)$.} 
\label{temp critique}
\end{center}
\end{figure}

\noindent We would like to  replace the parameters $\dc$ and $\dt$ that appear in the scale factor solution
by the temperatures $T_\pm$ at $t=0_\pm$. From the definition (\ref{rho}) (with $\dt$ replacing
$\d_T^2$) and the relation  (\ref{rhoT4}), using the fact
that $aT\equiv a_+T_+=a_-T_-$, one has
\be
\label{T+-}
T_\pm (\dc,\dt)=T_m {\sqrt{\delta_T^{\prime}}\over
  \sqrt{(1+\k')\left(1\pm\sqrt{1-\D}\right)}}\, ,
\ee
where the temperature $T_m$ has been defined in eq. (\ref{Tm}). On the
critical curve that is delimiting the domain (\ref{domain}), one has
$\dc=h(\dt)$. This implies $\D=1$ and thus $T_+=T_-$. It is shown
in the appendix that in this case, $\k'$ equals $\k'_{\D+}$ defined in eq. (\ref{k0+}). One can then deduce the critical
temperature $T_{c\mbox{\em\tiny T}}$ on the boundary of the domain (\ref{domain}) by $T_{c\mbox{\em\tiny T}}(\dt)\equiv T_\pm(\dt,h(\dt))$ (see fig. \ref{domain_fig}),
\be
T_{c\mbox{\em\tiny T}}(\dt)=T_m 
{\sqrt{{3\over 2}\delta_T^{\prime}}\over
  \sqrt{1+\sqrt{1-{3\over 4}\dt}}}\, ,
\ee
ploted on fig. \ref{temp critique}. 
The temperature on the critical curve can also be considered as a function of $\dc$. This can be
done by expressing $\dt=h^{-1}(\dc)$,
\be
\dt={4\over 3}-{4\over 3}\left( (A +B)^{1/3}+ (A-B)^{1/3}-{1\over 2}\right)^2\, ,
\ee
where
\be
A={1\over 8} +\dc \; , \quad B = \sqrt{\dc
  \left(\dc+{1\over 4}\right)}\;  , \quad \mbox{for}\quad\dc \le 1\,.
\ee
Defining $T_{c\c}(\dc)\equiv T_\pm(h^{-1}(\dc),\dc)$ (see fig. \ref{domain_fig}), one obtains
\be
T_{c\c}(\dc)=T_m 3^{1/4}\left({1- \left( ( A +B)^{1/3}+ (A-B)^{1/3}-{1\over 2}\right)^2
\over \left( (A +B)^{1/3}+ (A-B)^{1/3}+{1\over 2}\right)^2}\right)^{1/4}\, .
\ee

Since $T_+a_+=T_-a_-$, one  has   
\be 
\label{lD}
\D={4\over \left(T_+/T_-+T_-/T_+\right)^2}\, ,
\ee
a relation that generalizes eq. (\ref{Tm}) and the duality $T_+\leftrightarrow T_-$ .
Actually, for eqs. (\ref{lD}) and (\ref{Tm}) to have precisely the same form, one can introduce a mean temperature $T^{\prime
  2}_m\equiv T_+T_-$,
\be
\label{Tm'}
T'_m=T_m\left({\dt\over 4\k'^2+ 4\k'+\dt}\right)^{1/4}\, ,
\ee
where $T_m$ has been defined in eq. (\ref{Tm}). 
However, this definition is less useful than its counterpart in the pure thermal case. This is due to the fact that $T'_m$ 
depends on the point $(\dt,\dc)$ of the parameter space, while 
in the pure thermal case, $\dc=\k'=0$ so that the mean temperature $T_m$ is independent of the
remaining parameter $\dt$. 

We may express the evolution of the scale factor $a(\tt(t))$ in each branch with
$T_\pm$,
\be
a(\tt(t))={1\over\sqrt{2\l(1-\A)}}\sqrt{1+{T_-^2-T_+^2\over T_-^2+T_+^2}\cosh 
  (2\sqrt\l \,\tt(t))}\; , \quad t > 0 
\ee
and
\be
a(\tt(t))={1\over\sqrt{2\l(1-\A)}}\sqrt{1+{T_+^2-T_-^2\over T_+^4+T_-^2}\cosh 
  (2\sqrt\l\, \tt(t))}\; , \quad t < 0\, , 
\ee
where $\k'$ has been expressed in terms of $\A$ that involves  $T_\pm$ and $T_m$
\be
1+\k'= {1\over 1-\A}\quad\mbox{where}\quad
\A={1-{T^2_+T^2_-/T^4_m}\over \left(T_+/T_-+T_-/T_+\right)^2}\, .
\ee 
To obtain the above expression  for $\A$,  we have used eq. 
(\ref{Delta}) to express $\dt$ in terms of $\D$ and $\k'$. Then we combine the result with
 eqs. (\ref{T+-}) and (\ref{lD}).

Under the duality $T_+ \leftrightarrow T_-$,  the two cosmological solutions $a(\tt(t))$, for  $t>0$ and $t<0$ are interchanged :
\be
\left(\,T_+ \leftrightarrow T_-\right) \quad  \Longleftrightarrow \quad \left(\phantom{\dot \Phi}\!\!\!\!a(\tt(t))\; \mbox{for} \;  t >0   \leftrightarrow a(\tt(t))\; \mbox{for}\; t<0\right)\, . 
\ee 

For completeness  we display the constants $\N'$, $\ve'$ and
$\ve^{\prime\prime}$ appearing in eqs. (\ref{sol1}), (\ref{t1}) and
(\ref{exp3}),  in terms of $T_\pm$, 
\be
\N'={1\over \sqrt{\l(1-\A)}}\sqrt{{1-(T_+/T_-)^{2}\over
    1+(T_+/T_-)^{2} }}\; , \quad \ve'= {1 \over (T_-/T_+)^{2}-1}\; ,
\quad \ve^{\prime\prime}={1-{T^2_+T^2_-/T^4_m}\over
  (T_-/T_+)^2-(T_+/T_-)^2}\, .
\ee


\noindent{\em \large Parameterization with $\dt$ and $\dc$}

\noindent Let us determine explicit expressions of the scale factor in terms of $\dt$ and $\dc$, when $\D\le 1$. This will be useful in the next section, when we study the probability transition. Since the parameters $\dt$ and $\dc$ are chosen in the domain (\ref{domain}), the polynomial $\P$ in eq. (\ref{Pe}) has three real roots, (see appendix). Noting that $\P(-\l a_\pm)=0$, as can be checked with eq. (\ref{Delta}), the polynomial $\P$ is actually,
 \be
 \label{Pe}
\P(x)= (x-\k')(x+\l a_-^2)(x+\l a_+^2)\, .
\ee
Using Cardan's formulas, one finds 
\be
\label{k'}
\k'= {1\over 3}\left(\sqrt{4-3\dt}\cos\left({\theta\over 3}\right)-1\right)\, ,
\ee
\be
\label{a+-}
a^2_\pm = {1\over 3\l}\left(1-\sqrt{4-3\dt}\cos\left({\theta\pm2\pi\over3}\right)\right)\, ,
\ee
where
\be
\label{thet}
\theta= \arccos\left({16\dc+9\dt-8\over \left(4-3\dt\right)^{3/2}}\right)\, .
\ee
We note that all values of $\theta$ in $[0,\pi]$ can be reached. At the origin $(\dt,\dc)=(0,0)$ of the domain (\ref{domain}), one has $\theta=\pi$, while $\theta$ vanishes on the critical curve $\dc=h(\dt)$. The constant $\N'$ of eqs. (\ref{sol1}) and (\ref{exp3}) can be determined by expressing $\sqrt{1-\D}$ from the ratio $a^2_+/a^2_-$,
\be
\N'=\left({1-{3\over 4}\dt\over 3}\right)^{1/4}\sqrt{{2\over \l}\sin\left({\theta\over 3}\right)}\, .
\ee
The relations $a_-^2=\N^{\prime 2}\ve'$ and $\k'=\l\N^{\prime 2} \ve^{\prime \prime}$ give then
\be
\label{eps}
\ve'={1-\sqrt{4-3\dt}\cos\left({\theta-2\pi\over3}\right)\over \sqrt{3}\sqrt{4-3\dt}\sin\left({\theta\over 3}\right)}\; , \quad \ve^{\prime\prime}={\sqrt{4-3\dt}\cos\left({\theta\over3}\right)-1\over \sqrt{3}\sqrt{4-3\dt}\sin\left({\theta\over 3}\right)}\, .
\ee
For completeness, the temperatures at the transition are also found to be
\be
\label{T-}
T_\pm= T_m{\sqrt{{3\over 2}\delta_T^{\prime}}\over \sqrt{1-\sqrt{4-3\dt}\cos\left({\theta\pm 2\pi\over 3}\right)}}\, .
\ee


\subsection{First order transition probability}
\label{pro}

When $\D\le 1$, the transition probability at $t=0$ between the big bang branch and the deformed de Sitter cosmology
can be estimated by computing  the Euclidean action. Changing of  integration variable $d\tau={da_E /\dot a_E}$ in the Euclidean analog of the action (\ref{Sgene}) gives, on shell, 
\be
\label{ac}
S_{E\mbox{\tiny \em{eff}}}= -{1\over \l} \int _{a_-}^{a_+}{da_E \over a_E }  \left( \sqrt{\P(-\l a_E^2)}+{4\over 27}{\l^2\dc\over  \sqrt{\P(-\l a_E^2)}}\right)\, ,
\ee
where $\P$ is given in eq. (\ref{Pe}) and enters in the Euclidean equation of motion  
\be 
\left( {{\dot a}_E \over a_E } \right)^2= {N_E^2\over a_E^6} {\P(-\l a_E^2)\over \l^2}
\ee
we have used. Clearly, $S_{E\mbox{\tiny \em{eff}}}$ is negative. To explicitly  evaluate it, we prefer considering its integral form on $\taut$ and use eq. (\ref{Gtt}) in Euclidean time  to be on shell,
\be
\label{action2}
S_{E\mbox{\tiny \em{eff}}} =-\int_{-{\pi\over 2\sqrt{\l}}}^0 d\taut \, N_Ea_E^3\left( {1\over a_E^2}-\l-{\dt\over 4\l a_E^4}\right) \, .
\ee
In this expression, $a_E(\taut)$ and $N_E(\taut)\equiv d\tau / d\taut$ are referring to eq. (\ref{aE}). The result can be expressed in terms of complete elliptic integrals of the first kind, $K(k)$, and second kind, $E(k)$,
\be
S_{E\mbox{\tiny \em{eff}}}=-{1\over 3^{5/4}\l}\left(4-3\dt\right)^{1/4}\sqrt{\sin\left({\theta+\pi\over 3}\right)}\left( E(k)- {\sqrt{4-3\dt}\cos\left({\theta\over 3}\right)-1+{3\over 2}\dt \over \sqrt{3}\sqrt{4-3\dt}\sin\left({\theta +\pi\over 3}\right)} K(k)\right)\, ,
\ee
where 
\be
k=\sqrt{{\sin\left({\theta \over 3}\right)\over \sin\left({\theta +\pi\over 3}\right)}}\, .
\ee
To have a better intuition of the behavior of this action as a function of the parameters $\dt$ and $\dc$, we concentrate on the boundary of the domain (\ref{domain}) :\\
$\bullet$ On the side $\dc=0$ corresponding to the pure thermal case, the action becomes
\be
\label{acT}
S_{E\mbox{\tiny \em{eff}}}=-{1\over 3\l}\sqrt{{1+\sqrt{1-\dt}\over 2}}\left( E(k)-\left( 1-\sqrt{1-\dt}\right) K(k)\right)\, , 
\ee
where
\be 
k= {2(1-\dt)^{1/4}\over \sqrt{1+\sqrt{1-\dt}}}\, .
\ee
To derive these expressions, one can use the fact that $\k'$ in eq. (\ref{k'}) vanishes and is thus giving rise to an identity satisfied by $\theta$. Eq. (\ref{acT}) reproduces the result of \cite{Brustein:2005yn}\footnote{Up to minor misprint errors in \cite{Brustein:2005yn}.}.
At the origin $(\dt,\dc)=(0,0)$, one has $k=1$ and $E(1)=1$, so that  the result $-1/3\l$ of the pure de Sitter case is recovered. At $(\dt,\dc)=(1,0)$, $k=0$ and $E(0)=K(0)=\pi/2$, so that the action vanishes. This is consistent with eq. (\ref{ac}) since $a_+=a_-$ and $\dc$ vanishes.\\
$\bullet$ On the critical curve $\dc=h(\dt)$, the action is taking the form
\be
\label{aca}
 S_{E\mbox{\tiny \em{eff}}}=-{\pi\over 9\l} {\left(1+\sqrt{1-{3\over 4}\dt}\right)\left(\sqrt{1-{3\over 4}\dt}-1/2\right)\over \left(1-{3\over 4}\dt\right)^{1/4}}\, .
 \ee
If it is explicitly negative for $\dt\le 1$, it vanishes for $\dt=1$ only. This shows that even if $a_+\equiv a_-$ on the critical curve, the term proportional to $\dc$ in the integrand of eq. (\ref{ac}) contributes, due to the fact that the denominator vanishes. The probability $p$ refers in this case to the transition (after an infinite time) from the big bang branch  (\ref{inf}) to the inflationary one (\ref{sup}). The quantity $(1-p)$ refers to the transition (after an infinite time) form the big bang branch  (\ref{inf}) to its time reversal \ie describing a big crunch. This is the case since considering the solutions (\ref{sol1})  and (\ref{exp3}) of the case $\D<1$, and performing shifts on the origins of times, give rise to   (\ref{inf}) and (\ref{sup}) in the limit $\D\to1_-$.\\
$\bullet$ The side $\dt=0$ is corresponding to the pure moduli deformation of the de Sitter solution. From eq. (\ref{aca}), one finds that at the corner $(\dt,\dc)=(0,1)$ of the domain (\ref{domain}), the action is taking the value $-\pi/9\l$. Since this result is close to the $-1/3\l$ at the origin of the $(\dt,\dc)$-plane, the dependence of the action on $\dc$ seems to be mild along this axis. 

\begin{figure}[h!]
\begin{center}
\includegraphics[height=9cm]{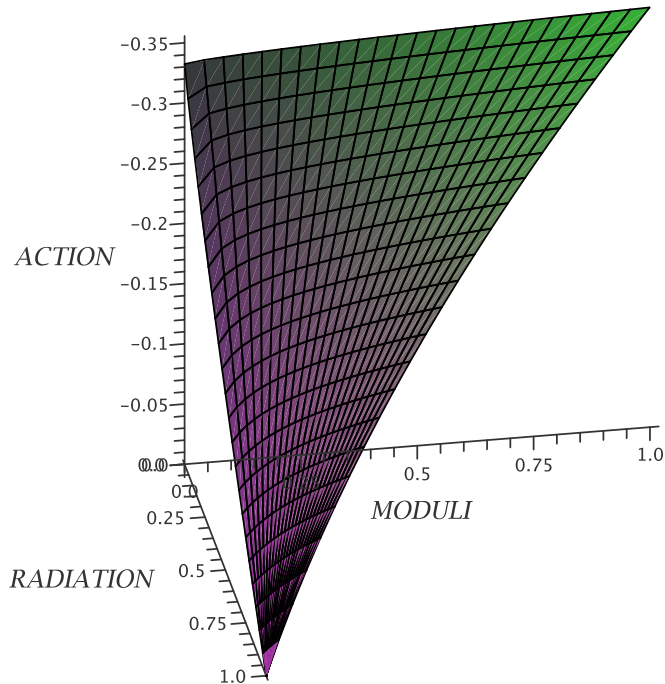}
\caption{\footnotesize \em Euclidean action $S_{E\mbox{\tiny
      \em{eff}}}$ as a function of the radiation parameter $\dt$ and the moduli
  deformation $\dc$. The variable $(\dt,\dc)$ spans the
  domain (\ref{domain}). The dependence on $\dc$ is weak, while the
  absolute value of $S_{E\mbox{\tiny \em{eff}}}$  decreases
  when $\dt$ is switched on. (We have chosen $\l=1$ on this plot.)}
\label{action3D}
\end{center}
\end{figure}

This remark happens to remain true for arbitrary fixed $\dt$. To
understand this, one can note in eq. (\ref{action2}) that the
$\dc/a^6$ contribution of $\dot \chi_i^2$ in the action
$S_{E\mbox{\tiny \em{eff}}}$ has disappeared on shell. Thus, the only
dependence on $\dc$ occurs through the mild deformation
$\ve^{\prime\prime}$ appearing in $N_E$. This is confirmed on the
3-dimensional plot of fig. \ref{action3D}, where $S_{E\mbox{\tiny \em{eff}}}$
appears almost constant when $\dc$ varies from $0$ to $h(\dt$), at
fixed $\dt$.

{F}rom fig. \ref{action3D}, one can also see that the transition amplitude $p \propto
e^{-2S_{E\mbox{\tiny \em{eff}}}}$ is larger and larger when $\dt$ decreases
from $h^{-1}(\dc)$ to $0$, at fixed $\dc$. However, the magnitude of $p$
highly depends on $\l$. For instance, $p$ may not be large even for small
$\dt$, due to the $1/\l$-dependance of $S_{E\mbox{\tiny \em{eff}}}$, for
large $\l$.  

At fixed $\l$ and $\dt$, switching on $\dc$ increases $\ve'$ (see
eqs. (\ref{eps}) and (\ref{thet})), and thus $2|t_i|$ (see  
eq. (\ref{ti})), the range of time along the big bang / big crunch
evolution. From eq. (\ref{T-}),  larger and larger $\dc$ imply also lower
and lower temperature $T_-$. As a result, increasing $\dc$ makes this
cosmological branch more and more viable. 

Similarly, at fixed $\l$ and $\dc\neq 0$, switching on $\dt$ has the effect
to 
increase $2 |t_i|$.  However, $T_-$ varies from $0$ to $T_{c\chi}$ , with
the risk for the universe to be too hot. The case $\dc =0$ has a different
behavior. When $\dt$ grows up from $0$ to $1$, $2 |t_i|$ increases as well
but the temperature $T_-$ varies from $+\infty$ to $T_m$.\footnote{Note
  that $T_- \to +\infty$ when $\dc\equiv 0$ and $\dt\to 0$, while
  $T_-\equiv 0$ when $\dt\equiv 0$ and $\dc\ne 0$. This  
ambiguity in the definition of $T_-$ in the pure de Sitter universe is due
to the fact that 
taking $\dc=0$ (\ie $\k'=0$) or $\dt=0$ are operations that do not commute 
(see eq. (\ref{Tm'})).} Along this axis, $\dt$ is thus making the big bang / 
big crunch branch more viable. 

Finally, when we reach the limit case corresponding to $(\dt,\dc)$ sitting on the critical curve, after the universe is born with a big bang, it converges quickly toward a static state for a very long time, \ie an $S^3$ of constant radius
\be
a\equiv a_\pm=a_0=\sqrt{{1+\sqrt{1-{3\over 4}\dt}\over 3\l}}\, ,
\ee
where $a_0$ has already been defined in eq. (\ref{Static}). However, it finally enters in an inflationary phase or dies in a big crunch. Also, one can note that in the case $\D>1$, the big bang evolution is longer and longer when $\D\to 1_+$, since $t_i\to -\infty$, (see eqs. (\ref{SOL2}) and (\ref{tbb})), but the final state on the universe is always inflationary.
 
\section{Conclusions}

In this work, we analyze some aspects concerning inflationary
solutions and possible transitions between different cosmological behaviors.
Our starting point is a bare de Sitter background deformed by the presence
of a thermal bath and the motion of moduli fields. The thermal bath
summarizes the effects of  light degrees of freedom in a consistent way and
shows up as a $C_R/a^4$ correction in the MSS-action. The moduli contribution
gives a $C_M/a^6$ correction to the energy density.

A systematic analysis of the system involves an almost triangular domain 
in the deformation parameter space $(\dt, \dc)$. We solve the gravitational equations in all cases and find that a first order phase transition can occur inside this domain, while a second order one arises outside. 

In the second case, a smooth transition always occurs between the big bang cosmological evolution (as soon as $T(t)<T_H$) and the inflationary behavior. In the first case, we calculate the transition probability between the big bang cosmological evolution
(where $T_-\le T(t)<T_H$) and the inflationary branch (where $T(t)\le T_+$).
The two solutions are connected into each other via a gravitational
instanton allowing a double analytic continuation. 
We find, the existence of  a temperature duality $T_+ \leftrightarrow T_-$ 
that interchanges the two solutions. Inside the domain of the parameter space, the origin $(\dt,\dc)\simeq(0, 0)$ corresponds to the pure de Sitter case where the big-bang
branch disappears and the instanton connects the inflationary universe
to ``nothing''. In the two other extreme cases,  $i)$ the pure thermal deformation 
  $(\dt,\dc)\simeq(1, 0)$ and $ii)$ the pure moduli deformation $(\dt,\dc)\simeq(0, 1)$, the big bang branch remains almost static for a very long period of time. However, the probability transition is minimal in case $i)$ and maximal in case $ii)$. Thus, the late future of these evolutions is most probably contracting (till a big crunch occurs) in case $i)$, and
inflationary in case $ii)$.

The above analysis and results can be applied to more complex systems 
where the radiative and temperature corrections are effectively taking the form of cosmological, curvature and radiation terms during the time evolution \cite{KP}. These
richer models occur in string effective no-scale supergravities \ie associated to $N=1$ string compactifications  where $N=1$ supersymmetry is spontaneously broken. First and second order phase transitions are again involved \cite{KP}. The stringy origin of these models is giving us the
possibility to go beyond the field theory approximation. In particular, it is possible to study the 
big bang cosmological evolution above the Hagedorn temperature \cite{KPTT} 
as well as ``the stringy analog of the wave function of the universe" \cite{KTT}.   
 

\section*{Acknowledgements}

We are grateful to Constantin Bachas,  Ioannis Bakas, Gary Gibbons and John Iliopoulos for  discussions. \\
\noindent
The work of C.K. and H.P. is partially supported by the EU contract MRTN-CT-2004-005104 and the ANR (CNRS-USAR) contract  05-BLAN-0079-01. C.K. is also supported by the UE contract MRTN-CT-2004-512194, while H.P. is supported by the UE contracts MRTN-CT-2004-503369 and MEXT-CT-2003-509661, INTAS grant 03-51-6346, and CNRS PICS 2530 and 3059.


\vspace{.2cm}
\begin{center}
{\Large\bf Appendix}
\end{center}
\renewcommand{\theequation}{A.\arabic{equation}}
\renewcommand{\thesection}{A.}
\setcounter{equation}{0}

\noindent We would like to find in this appendix the phase diagram in the
$(\dt,\dc)$-plane that is delimiting when the solutions (\ref{sol1}) and (\ref{exp3}) for the scale factor are valid, and when it is instead the evolution (\ref{SOL2}) that is relevant. The two first solutions exist when $\D\le1$, while the last one occurs when $\D\ge 1$.   

First of all, we note that the polynomial $\P$ defined in
eq. (\ref{d3}) has three roots, whose product is  $4\d_\c^2/27$. If
two of them are complex conjugate, the third one is thus real
positive. If on the contrary they are all real, one or three of them
must be positive. However, since the sum of these real roots is
$-1$, only one is positive. In any case, we always have a single
real positive root, $\k'$, and eventually two other real negative ones.  

For $\dt>4/3$, the derivative $\P'$ of the polynomial defined in eq. (\ref{d3}) is everywhere positive, so that $\P$ has a single root. When 
\be
\label{cons}
\dt\le {4\over 3}\, ,
\ee
the derivative $\P'$ admits two roots, 
\be
\k'_\pm = -{1\over 3} \left( 1\mp \sqrt{1-{3\over 4}\dt}\right)\, .
\ee
One always has $\P(\k'_+)\le 0$, while $\P(\k'_-)\ge 0$  (so that $\P$ has 3 real roots) if and only if $\dc\le h(\dt)$, where $h(\dt)$ is defined in
eq. (\ref{h}). Actually, this condition implies in particular that
$h(\dt)\ge 0$, which is equivalent to $\dt\le 1$. As a
conclusion, the polynomial $\P$ has three real roots, (one positive and two negative), if the
inequalities (\ref{domain}) are satisfied, while it has only one real root in all other
cases.

Let us determine when $\D\le 1$. This condition can be translated into a degree two inequality for $\k'$ that admits solutions if the condition (\ref{cons}) is satisfied. In that case, $\D\le 1$ is equivalent to having  
\be 
\k'_{\D-}\le \k'\le \k'_{\D+}\, ,
\ee
where
\be
\label{k0+}
\k'_{\D\pm}={1\over 3}\left( \pm \sqrt{4-3\dt}-1\right)\, .
\ee
Since having $\k'_{\D+}< 0$ is equivalent to satisfying $\dt>1$, the single real root $\k'$
that $\P$ has in this case is outside the range $[\k'_{\D-},\k'_{\D+}]$, since it is positive. Thus, $\dt>1$ implies $\D>1$. 

Let us then concentrate on the case $\dt\le 1$. We note that when the parameters $\dc$ and $\dt$ are chosen on
the critical curve defined by $\dc=h(\dt)$, one has
$\P(\k'_{\D+})=0$. This means that the single real positive root $\k'$ of
$\P$ is then precisely $\k'_{\D+}$. In that case, we have $\D=1$. If instead we take
$\dc>h(\dt)$, the only real root of $\P$ is positive and satisfies
$\k'>\k'_{\D+}$, due to the fact that $\P'(x)> 0$ for any
$x>0$. We thus have $\D>1$. On the contrary, if $\dc<
h(\dt)$, the real root of $\P$ which is positive satisfies
$\k'<\k'_{\D+}$ for the same reason, so that we have $\D<1$. 

As a conclusion, the solutions (\ref{sol1}) and (\ref{exp3}) corresponding to the case $\D\le 1$ arise when the point $(\dt,\dc)$ belongs to the interior of the domain
(\ref{domain}) of fig. \ref{domain_fig}. Outside this domain, the evolution (\ref{SOL2}) associated to $\D\ge 1$ is the correct one. On the critical curve $\dc=h(\dt)$, one has $\D=1$.




\begin{thebibliography}{99}

\bibitem{Witten:1981gj}
  E.~Witten,
  ``Instability of the Kaluza-Klein vacuum,''
  Nucl.\ Phys.\ B {\bf 195} (1982) 481.

\bibitem{Vilenkin:1982de}
  A.~Vilenkin,
  ``Creation of universes from nothing,''
  Phys.\ Lett.\ B {\bf 117} (1982) 25.
  
\bibitem{Vilenkin:1983xq}
  A.~Vilenkin,
  ``The birth of inflationary universes,''
  Phys.\ Rev.\ D {\bf 27} (1983) 2848.
   
\bibitem{Hartle:1983ai}
  J.~B.~Hartle and S.~W.~Hawking,
  ``Wave function of the universe,''
  Phys.\ Rev.\ D {\bf 28} (1983) 2960.

\bibitem{Hawking:1984hk}
  S.~W.~Hawking,
  ``The cosmological constant is probably zero,''
  Phys.\ Lett.\  B {\bf 134} (1984) 403.

\bibitem{Sarangi:2005cs}
  S.~Sarangi and S.~H.~Tye,
  ``The boundedness of Euclidean gravity and the wavefunction of the universe,''
  arXiv:hep-th/0505104.
    
\bibitem{Sarangi:2006eb}
  S.~Sarangi and S.~H.~Tye,
  ``A note on the quantum creation of universes,''
  arXiv:hep-th/0603237.

\bibitem{Watson:2006px}
  S.~Watson, M.~J.~Perry, G.~L.~Kane and F.~C.~Adams,
  ``Inflation without inflaton(s),''
  arXiv:hep-th/0610054.
  
\bibitem{Brout:1989pw}
  R.~Brout and P.~Spindel,
  ``Tunneling in cosmology and isothermal inflation,''
  Nucl.\ Phys.\  B {\bf 348} (1991) 405.
  
\bibitem{Brustein:2005yn}
  R.~Brustein and S.~P.~de Alwis,
  ``The landscape of string theory and the wave function of the universe,''
  Phys.\ Rev.\ D {\bf 73} (2006) 046009
  [arXiv:hep-th/0511093].
  
\bibitem{BouhmadiLopez:2006pf}
  M.~Bouhmadi-Lopez and P.~Vargas Moniz,
  ``Quantisation of parameters and the string landscape problem,''
  JCAP {\bf 0705} (2007) 005
  [arXiv:hep-th/0612149].

\bibitem{KP}
  C.~Kounnas and H.~Partouche,
  ``Inflationary de Sitter solutions from superstrings,''
  arXiv:0706.0728 [hep-th].

\bibitem{Hawking:1983hj}
  S.~W.~Hawking,
  ``The quantum state of the universe,''
  Nucl.\ Phys.\  B {\bf 239} (1984) 257.

\bibitem{Harrison}
  E.~R.~Harrison,
  ``Classification of uniform cosmological models,''
  Mon.\ Not.\ R.\ Astron.\ Soc.\ {\bf 137} (1967) 69.
  
\bibitem{Kamenshchik:1995ib}
  A.~Y.~Kamenshchik, I.~M.~Khalatnikov and A.~V.~Toporensky,
  ``Nonminimally coupled complex scalar field in classical and quantum
  cosmology,''
  Phys.\ Lett.\ B {\bf 357} (1995) 36
  [arXiv:gr-qc/9508034].

\bibitem{Barvinsky:2006uh}
  A.~O.~Barvinsky and A.~Y.~Kamenshchik,
  ``Cosmological landscape from nothing: Some like it hot,''
  JCAP {\bf 0609} (2006) 014
  [arXiv:hep-th/0605132].
  
\bibitem{Firouzjahi:2004mx}
  H.~Firouzjahi, S.~Sarangi and S.~H.~H.~Tye,
  ``Spontaneous creation of inflationary universes and the cosmic  landscape,''
  JHEP {\bf 0409} (2004) 060
  [arXiv:hep-th/0406107].

\bibitem{KPTT} 
In preparation.

\bibitem{KTT} 
 C.~Kounnas, N.~Toumbas and J.~Troost,
 ``A wave-function for stringy universes,''
 arXiv:0704.1996 [hep-th].
    
\end{thebibliography}
\end{document}